\pdfoutput=1
\documentclass[10pt,a4paper,twocolumn]{article}
\usepackage[T1]{fontenc}
\usepackage[utf8]{inputenc}
\usepackage{lmodern}
\usepackage[margin=1.8cm,columnsep=7mm]{geometry}
\usepackage{microtype}
\usepackage{amsmath,amssymb,amsthm,mathtools}
\usepackage{graphicx}
\usepackage{booktabs}
\usepackage{stfloats}
\usepackage[font=small,labelfont=bf,skip=4pt]{caption}
\usepackage{xcolor}
\definecolor{colAB}{HTML}{4477AA}\definecolor{colABC}{HTML}{44AA99}\definecolor{colDE}{HTML}{EE7733}\definecolor{colRej}{HTML}{888888}% palette shared with Fig. 2
\usepackage{tikz}
\usepackage{rotating}
\usepackage[authoryear,round]{natbib}
\bibpunct{(}{)}{;}{a}{}{:}
\usepackage[hidelinks]{hyperref}
\usepackage{orcidlink}
\newcommand{\rr}[1]{{\raggedright #1\par}}
\newenvironment{ruledtabular}{\par\centering\small\setlength{\tabcolsep}{4pt}}{\par\smallskip}

\theoremstyle{remark}

\title{Testing stemmata against the evidence:\\ Exact optimisation, evidence maps and Bédier's question}
\author{Christophe Jurczak\,\orcidlink{0009-0001-8204-7333}\\[2pt]\small Metamorphoses of Civilization Collection, Dallas, TX, USA\\\small\texttt{cj@metamorphoses-collection.org}}
\date{}
\usepackage{float}
\begin{document}
\twocolumn[{\csname @twocolumnfalse\endcsname
\maketitle\vspace{-1em}
\begin{abstract}
Critical editions rest on stemmata, the family trees of a text's manuscripts, whose reliability has been doubted since Bédier (1928). Which branches of a manuscript stemma are established by the readings, and which require further editorial judgement? We formulate stemma reconstruction as a maximum-weight independent set: shared readings propose manuscript families, weighted by their support, and exact optimisation selects the heaviest compatible set, including jointly completable families from fragmentary witnesses. The result is a tree with an evidence map identifying the readings supporting and challenging each branch and its stability under resampling.

On an artificial tradition of known history, exact optima perform comparably to leading reconstruction methods. For the passion of Donatus of Arezzo, the evidence map audits the editor's stemma: most of his families are firmly attested, a few rest on a handful of readings, and the family at the top of his stemma has no supporting reading in the apparatus. For seven copies of an unedited prologue to the passion of St Lawrence, the method constructs a new stemma. Across fourteen texts of the \emph{Monumenta Germaniae Historica}, the editors' own apparatus supports four top divisions, contradicts four and leaves six undecided. In Bédier's \emph{Lai de l'Ombre}, the readings establish a stable unrooted tree, while its root depends on which readings are judged original.

The conflict graphs also admit a physical implementation: small unweighted instances were sampled on Pasqal's Ruby neutral-atom quantum processor, with weighted optimisation demonstrated in emulation. The method makes the evidence behind a stemma auditable.
\end{abstract}
\noindent\textbf{Keywords:} stemmatology; textual criticism; critical apparatus; \emph{Monumenta Germaniae Historica}; Bédier; quantum computing.
\vspace{1.2em}}]

%=====================================================================
\section{Introduction}
%=====================================================================

\begin{figure*}[t]
\centering
\begin{minipage}[b]{0.385\textwidth}\centering\includegraphics[height=4.1cm]{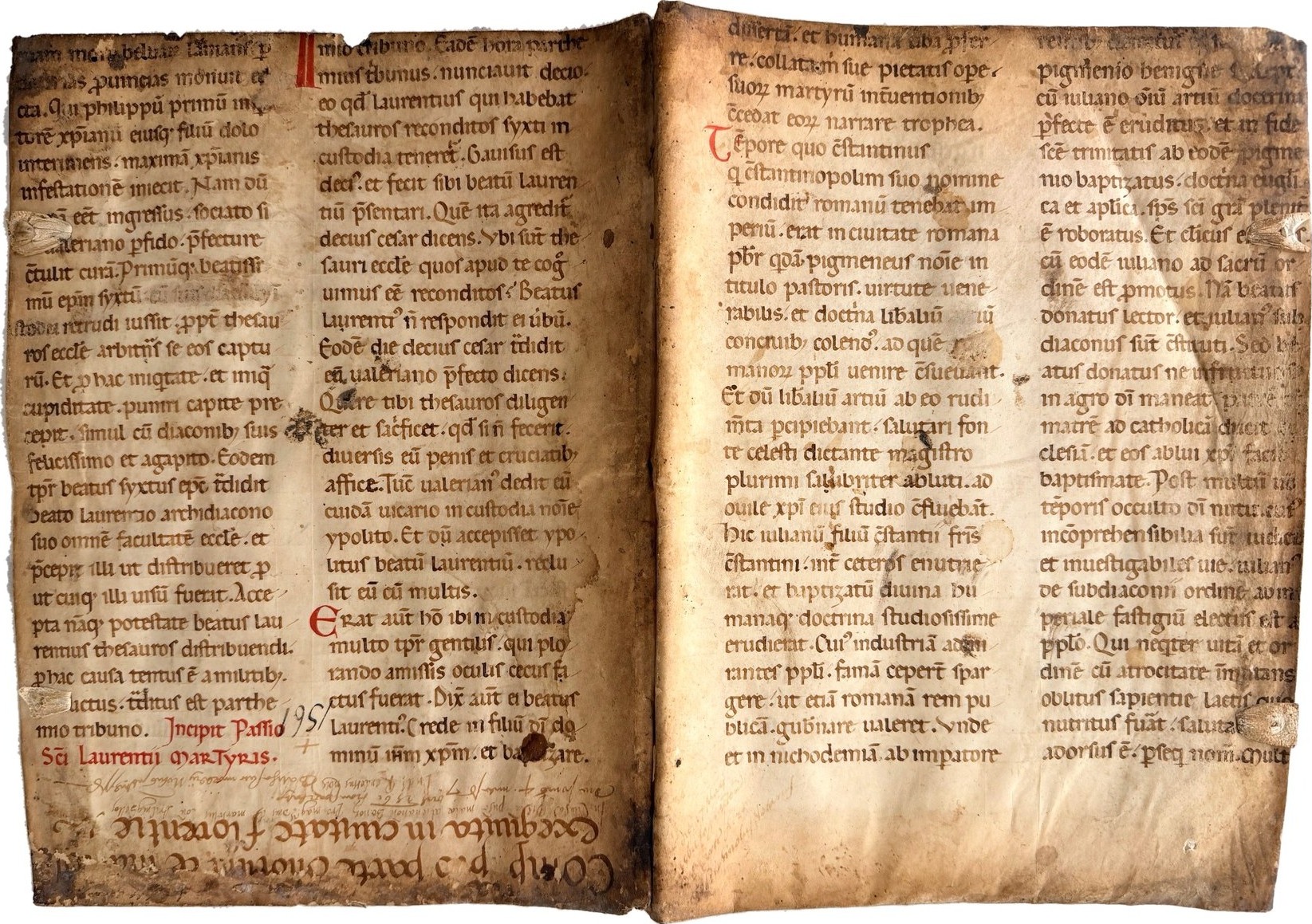}\\[-1pt]{\small (a)}\end{minipage}\hfill
\begin{minipage}[b]{0.385\textwidth}\centering\includegraphics[height=4.1cm]{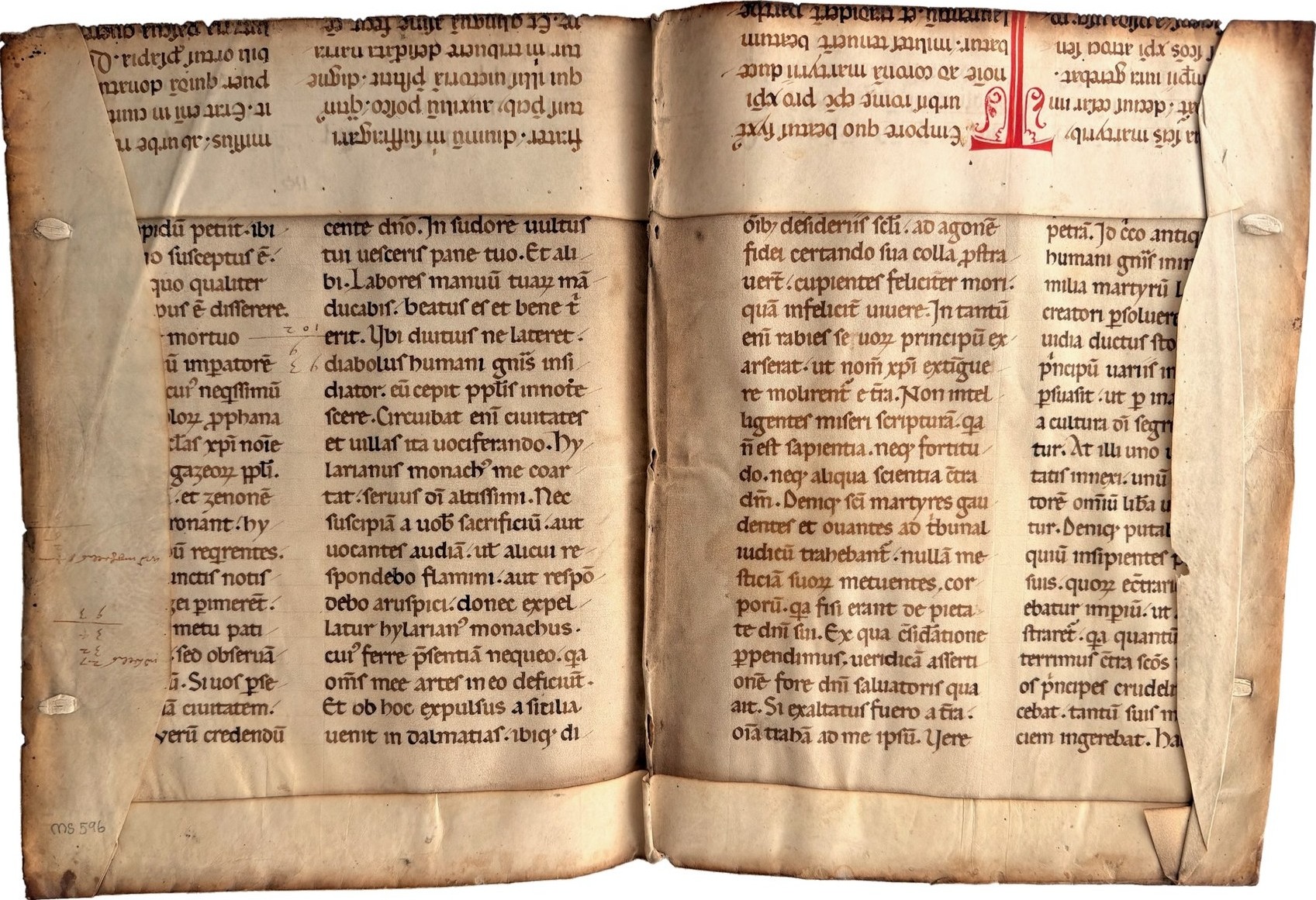}\\[-1pt]{\small (b)}\end{minipage}\hfill
\begin{minipage}[b]{0.20\textwidth}\centering\includegraphics[height=4.1cm]{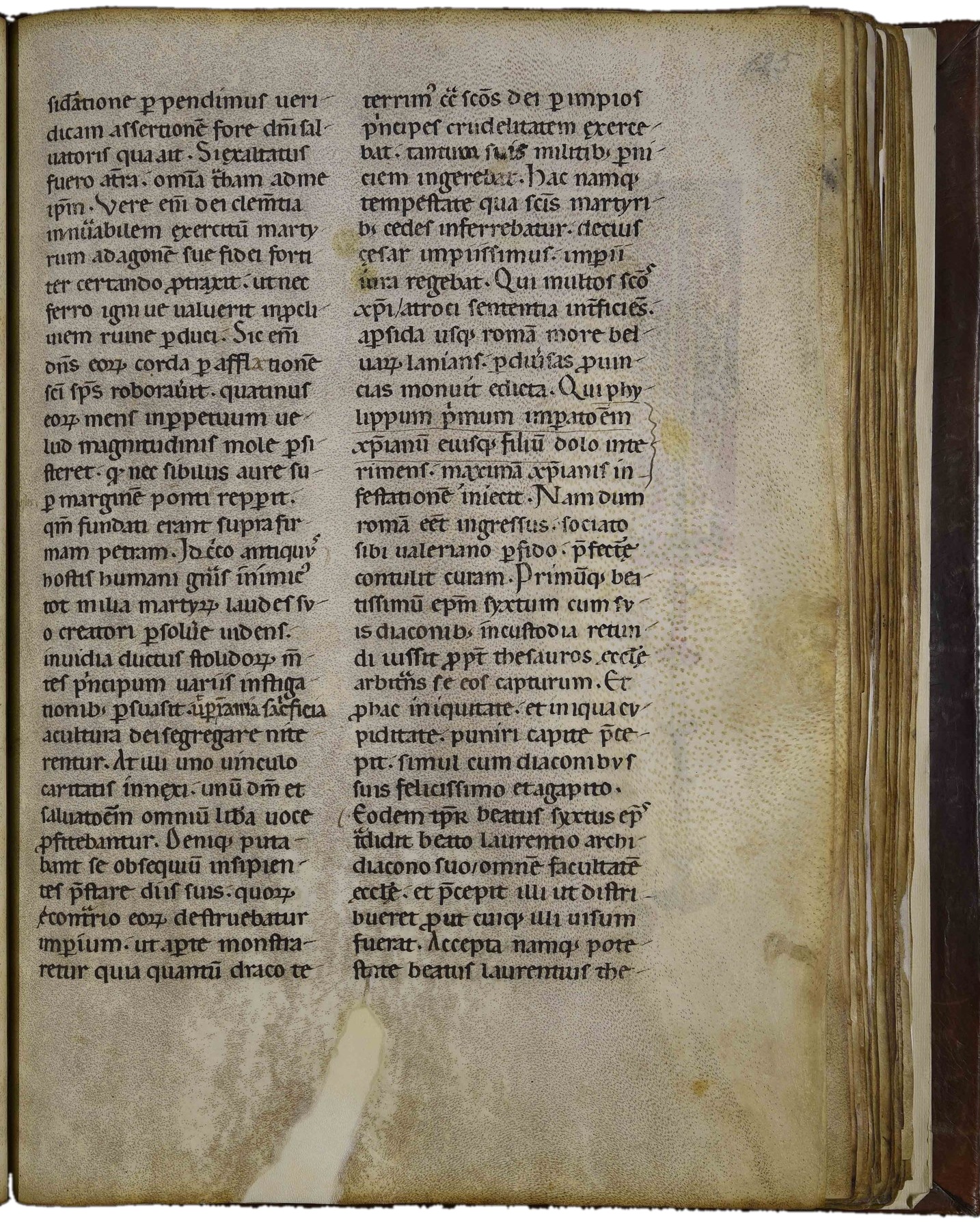}\\[-1pt]{\small (c)}\end{minipage}
\caption{\label{fig:plates}Bifolium L (MC-C-077), second half of the twelfth century. (a)~Outer face (fols 2v, 1r), worn by use as a file cover in 1561. (b)~Inner face (fols 1v, 2r): end of the passion of Donatus of Arezzo and the prologue \emph{In divinis laudibus}. (c)~Trento, Biblioteca comunale, BCT1-1566 (passional T), p.~125, in the prologue. Images: (a, b) Metamorphoses of Civilization Collection; (c) Biblioteca Digitale Trentina, public domain. L and T are the two closest of the seven witnesses of the prologue (BHL 4759): the method of this paper pairs them in 98\% of bootstrap replicates.}
\end{figure*}

This paper starts from a singular object. A bifolium of the second half of the twelfth century, reused in 1561 as the cover of a Florentine lawsuit file (Fig.~\ref{fig:plates}a, b), carries the end of the passion of Donatus of Arezzo (\emph{Bibliotheca Hagiographica Latina}, BHL 2294) and part of a prologue to the passion of St Lawrence, \emph{In divinis laudibus} (BHL 4759). Placing it among the other copies requires the stemma of each text. For Donatus a modern critical edition exists, built on 54 manuscripts with a full apparatus~\citep{Licciardello2018}; the prologue had no edition, and a reading text from its seven copies is published online with a translation~\citep{MCPassional}. Both raise the same question: which branches of a stemma does the evidence actually establish, and which rest on the editor's judgement? The two traditions are test cases of the method presented here, one checked against a published stemma, the other built from scratch. The method is tested first on an artificial tradition of known history, then on these two, and finally on a whole series of nineteenth- and early twentieth-century editions.

That question goes to the foundations of textual criticism. Every ancient or medieval text reaches us through copies of copies. What survives is a set of manuscripts, the \emph{witnesses}, each of which inherited the text of its exemplar together with the exemplar's mistakes, and added mistakes of its own. The editor works back from the witnesses to the text of their latest common ancestor, the \emph{archetype}, and the instrument of that work is the \emph{stemma codicum}: a tree whose leaves are the surviving manuscripts, whose internal nodes are lost intermediaries, and whose edges record copying~\citep{Maas1958,Timpanaro2005}. Once the stemma is known it guides the choice between readings, since a reading shared by two independent branches is more likely original than one found in a single branch. Textual criticism is both an art and a science~\citep[5]{Roelli2020}; this paper concerns the science, the evidence on which the editor's judgement rests.

Two answers to the question of how a stemma is built have shaped the field. The genealogical method associated with Karl Lachmann, the nineteenth-century German philologist, assembled largely by his predecessors~\citep{Timpanaro2005} and set out as a system by the classical scholar Paul Maas in 1927~\citep{Maas1958}, groups manuscripts by their shared errors, which orients the tree from the archetype down. Dom Henri Quentin, a Benedictine monk editing the Vulgate, classified manuscripts in 1926 by all their variants, without first deciding which readings are errors, and oriented the resulting network only afterwards~\citep{Quentin1926}; the computer scientist Gian Piero Zarri implemented his method on a computer in the 1970s~\citep{Zarri1976}. Shared readings imply a common origin but not where it lies: they determine an unrooted tree, and only a judgement on which readings are innovations gives it a root. Since the French medievalist Joseph Bédier, editor of the \emph{Chanson de Roland}, observed in 1928 that almost all the stemmata he had collected from editions of French texts split into two branches at the top~\citep{Bedier1928}, the method has lived with a suspicion of subjectivity~\citep{Timpanaro2005}, which models of copying and loss have only partly answered~\citep{Weitzman1987,GuidiTrovato2004,Camps2025}.

Computer methods came to stemmatology from evolutionary biology: parsimony, distance methods and Bayesian inference have been applied to the \emph{Canterbury Tales}~\citep{Barbrook1998}, to artificial traditions~\citep{Spencer2004,RoosHeikkila2009} and to real ones~\citep{McCollumTurnbull2024,Turnbull2024}, alongside methods designed for texts~\citep{RoosHeikkila2009,RoosZou2011,AndrewsMace2013,CampsCafiero2015} and databases of published stemmata~\citep{OpenStemmata2021,Bizais2026}; their history is surveyed by~\citet{Hoenen2020history}. The \emph{Handbook of Stemmatology} notes three limits that matter here~\citep[310, 347, 353]{Roelli2020}: the searches are heuristic, most programs return bifurcating trees by construction, and missing text and contamination, the copying of a text from more than one exemplar, remain unsolved.

We state the editor's own logic as an exact optimisation problem. If each shared reading proposes a family, a stemma is a heaviest collection of proposed families that do not cross: a maximum-weight independent set (MWIS) on the graph whose vertices are the families and whose edges join families that cross. In phylogenetics this is the \emph{compatibility} method~\citep{LeQuesne1969,Estabrook1976}; stemmatology has used the idea locally, to choose between incompatible groups~\citep[121]{Roelli2020}, to break conflicts one at a time~\citep[62--65]{Dearing1974} or to weight likely \emph{Leitfehler}~\citep{RoelliBachmann2010}. Here the compatible families are themselves the answer, chosen by a global optimum. The cladistic reading of the genealogical method goes back to~\citet{PlatnickCameron1977}; we know of no earlier use of a global compatibility optimum over a whole manuscript tradition. Exactness extends to fragmentary witnesses through a cutting-plane loop, since pairwise tests no longer suffice there~\citep{Steel1992,Shutters2013}; it is transparent, since every family comes with the readings that support it and the families that exclude its rivals; and on a neutral-atom processor~\citep{Henriet2020,Dalyac2024} the Rydberg blockade itself forbids two crossing families to be kept together (maximum compatibility has also been formulated as an MWIS and sampled on gate-based hardware for viral sequences;~\citealp{Ye2026}). The maximum-weight independent set is the clearest of the problems that such processors solve natively, for conflict graphs that fit the geometry of the array, and it is the logic an editor applies to shared readings.

Exactness certifies that the stated objective is optimal; it does not guarantee a more accurate reconstruction, and on the benchmark with known history the scores of exact and heuristic methods overlap. Its value is that it audits a stemma: it separates the science, what the readings decide, from the art, what rests on the editor's judgement (the root, near-equal groupings, contamination), and shows which readings each decision turns on.

The paper makes four contributions.
(i) Stemma reconstruction as exact weighted compatibility, with fragmentary witnesses treated by joint completion of partial splits and cutting planes (Sec.~\ref{sec:method}).
(ii) The \emph{evidence map} of a published stemma, the support, conflict and stability of each family with the loci behind them, tested on a benchmark with known history and on a 54-witness edition spot-checked against images of three witnesses, with fifteen generative language models compared under the same protocol (Sec.~\ref{sec:results}).
(iii) Five founding editions of the \emph{Monumenta Germaniae Historica} read back from their apparatus, a test across a whole series, with thresholds fixed in advance, of whether each editor's own apparatus supports his top division, and Bédier's own text as the limiting case (Sec.~\ref{sec:founding}).
(iv) Neutral-atom processors as physical solvers of the stemma: small instances run on a processor of up to 100 atoms, and the effect of per-atom weights shown in emulation (Sec.~\ref{sec:quantum}).
The mathematics behind the method (identifiability, error bounds, and the over-resolution that gives Bédier's suspicion a ground) is proved in full in the technical report that accompanies this paper~\citep{JurczakReport}, where every result is also checked by machine and verified in the proof assistant Lean.

%=====================================================================
\section{Method}
\label{sec:method}

This section states the method in the editor's terms. Formal definitions, complete proofs and the machine checks of every result are in the technical report.

\subsection{Collations, families and stemmata}
\label{sec:setting}

\begin{figure*}[t]
\centering
\includegraphics[width=\textwidth]{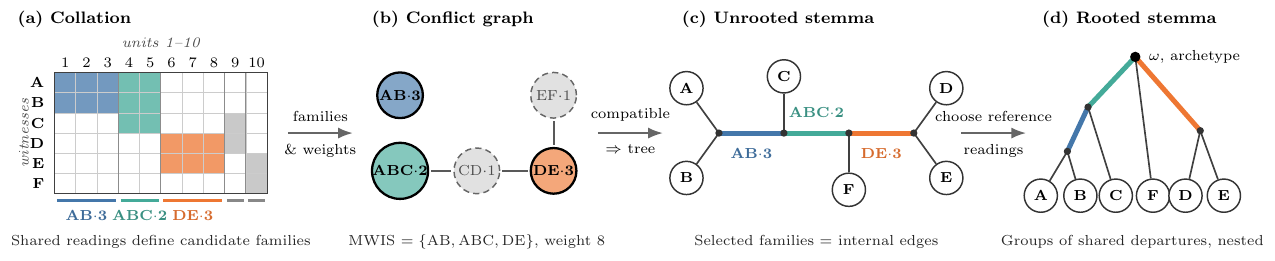}
\caption{Stemma reconstruction as a maximum-weight independent set, on six witnesses and ten units. (a)~Collation: within each column, coloured cells share one reading; each group is a candidate family, weighted by the number of units that propose it. (b)~Conflict graph: crossing families are joined; the heaviest set of non-adjacent vertices is $\{AB, ABC, DE\}$, of weight 8. (c)~The compatible families are the internal edges of a unique unrooted tree (Buneman's theorem, Sec.~\ref{sec:complete}). (d)~With the white readings as reference, the families nest and the tree is drawn from the archetype $\omega$ down; the top has three branches, ABC, DE and F.}
\label{fig:mwis}
\end{figure*}
%=====================================================================

A \emph{collation} records, for each place of variation (a \emph{unit}), which witnesses are legible there and what each of them reads. Each reading proposes a \emph{family}: the witnesses that share it, against the others. In the language of phylogenetics a family is a \emph{split} of the witnesses into two sides. A unit with several readings proposes several families, and the same family proposed at different units is counted once per unit (the two readings of a binary unit propose the same split, counted once); its \emph{weight} is the number of units that propose it. In the editor's vocabulary, the weight of a family is the number of places where its members share a reading.

A \emph{stemma} is drawn here as a tree whose leaves are the surviving witnesses and whose internal nodes are lost exemplars; cutting any internal edge divides the witnesses into a family. Such a tree records grouping, not direction of copying, so a surviving exemplar of other witnesses appears as a leaf beside them. It says which witnesses belong together, not which readings are original: that is decided by its \emph{root}, the place of the archetype, which the shared readings alone do not fix (Sec.~\ref{sec:complete}). Two families are \emph{compatible} when one of the four intersections of their sides is empty, and \emph{in conflict} when they cross: some manuscripts belong to both, some to the first only, some to the second only, and some to neither (the ``four-gamete'' test of phylogenetics); for families oriented away from a reference, compatible means disjoint or nested. The \emph{conflict graph} has one vertex for each proposed family, weighted by the number of units that propose it, and an edge between every two families that cross.

\subsection{From collation to stemma}
\label{sec:complete}
%=====================================================================

A classical theorem of Buneman links the two: a set of families is the set of branches of some stemma exactly when its families are pairwise compatible, and that stemma is then unique~\citep{Buneman1971,SempleSteel2003}. A stemma is thus the same thing as a set of pairwise compatible families. We take as best supported the stemma whose families are proposed by the most units; with a complete collation it is the heaviest set of families of which no two cross. On the conflict graph, with vertices $V$, edges $E$, weight $w(\sigma)$ for family $\sigma$, and $y_\sigma=1$ when family $\sigma$ is kept in the stemma and $0$ when it is left out, this is a maximum-weight independent set (MWIS):
\begin{equation}\label{eq:mwis}
\begin{aligned}
\max\ & \textstyle\sum_{\sigma\in V} w(\sigma)\, y_\sigma\\
\text{s.t.}\ & y_\sigma + y_\tau \le 1\quad \forall\{\sigma,\tau\}\in E,\qquad y\in\{0,1\}^V .
\end{aligned}
\end{equation}
Figure~\ref{fig:mwis} shows the steps on a toy tradition. For binary units this is the \emph{compatibility} method of numerical taxonomy~\citep{LeQuesne1969,Estabrook1976}; at a unit with three or more readings each reading's family votes separately, which differs from the compatibility of whole multistate characters. Exact algorithms for it exist. Cherry's clique-based method handles nearly two thousand bacterial genomes, data with far less conflict than a manuscript tradition, and already checks whether pairwise-compatible ambiguous splits can be resolved together, refining the solution when they cannot~\citep{Cherry2017}; Ye and colleagues formulate maximum compatibility as the enumeration of maximum-weight independent sets on the complement graph~\citep{Ye2026}. What is new here is the coding of a collation into families, the treatment of fragmentary witnesses within it, the evidence map, and their application to manuscript traditions. Since all weights are positive, the optimum keeps every proposed family that crosses none of those selected, and so tends to divide the witnesses into more groups than the true stemma has (Sec.~\ref{sec:conclusion}).

The tree so chosen is unrooted. To root it, the method follows Lachmann's rule that only shared innovations prove kinship: witnesses that keep the original reading are not thereby related. A reference text is fixed, the editor's lemma or the majority reading, and at each unit only the witnesses that depart from it together form a family. Two such families are compatible when they are disjoint or one contains the other, a compatible set of them is a tree that hangs from the reference, and the best-supported set is again the solution of~\eqref{eq:mwis}; formally, this is the unrooted problem with the reference text added as one more witness. Because the reference decides which readings count as departures, the rooted optimum need not be the unrooted tree with a root added: its families can differ. In the \emph{Lai} and in Thegan they do not, and only the orientation changes (Secs.~\ref{sec:lai} and~\ref{sec:thegan}). The tree is drawn from the archetype down, each family below the families that contain it (Fig.~\ref{fig:mwis}d); all rooted trees in this paper are drawn this way, with the reference stated. A family of all witnesses but one, such as B, M and P against E in de Boor's stemma of Priscus (Sec.~\ref{sec:ruby}), says nothing in the unrooted tree, where it only sets E apart, and becomes a branch once the tree is rooted. The root is therefore not read from the evidence but taken from the reference text, with whatever judgement went into it. This is where the editor's art enters, and Sec.~\ref{sec:mommsen} reports how much the results depend on it.

The definition of best supported carries assumptions that the method does not test. Every unit weighs the same, whatever the kind of variant. A unit counts for a family only if its reading set is exactly that family, so a pattern that needs one more change gives nothing. Units are treated as independent, so an omission or a rewritten sentence split into several units counts several times. The result is always a tree, so contamination shows only as families that drop out. And the root comes from outside the readings. Each can mislead; the sections below report where they matter.

In general the problem is hard: MWIS is NP-hard~\citep{Karp1972}, and remains so on conflict graphs of families, since every finite graph is the conflict graph of some set of splits~\citep{DaySankoff1986}. The difficulty grows with the number of witnesses, not of variants: with $n$ witnesses, a dynamic programme finds the optimum in $O(3^n)$ operations whatever the number of units. In practice Donatus, the largest tradition in this paper (54 witnesses, 1{,}425 candidate families), is solved exactly in seconds; larger traditions have not been tried.

%=====================================================================
\subsection{Fragmentary witnesses}
\label{sec:partial}
%=====================================================================

Incomplete witnesses are where texts differ most from the usual phylogenetic setting. When leaves are lost or lines illegible, a unit proposes a \emph{partial} family, defined only on the witnesses it observes. The natural move for an editor is to declare two partial families in conflict when they cross on the witnesses both observe, which gives a graph to which~\eqref{eq:mwis} still applies. That test is exact for pairs: two partial families that do not cross where both are observed can always be completed, by placing the unobserved witnesses, into full families that do not cross. It fails for triples: three partial families can be pairwise consistent and still admit no common completion. The smallest such triples need five witnesses; there are 200 of them on five witnesses and none on four (the phenomenon is known for quartets;~\citealp{Shutters2013}).

The pairwise graph is therefore the right starting point, and needs correction rather than replacement. We solve~\eqref{eq:mwis} on it, test whether the optimum can be jointly completed, and if not add a constraint that excludes the offending subset, repeating until the optimum is completable; the first completable optimum is then exact, by a standard cutting-plane argument, close in spirit to Cherry's treatment of ambiguous splits~\citep{Cherry2017}. The completability test is a satisfiability problem, solved with CP-SAT~\citep{ORTools}. A completable set of partial families is in general displayed by several trees, so the families selected need not fix the position of a fragmentary witness; we report the completion found, not all of them.

\section{Testing the method}
\label{sec:results}

\subsection{Benchmarks}
\label{sec:experiments}
\begin{table}[t]
\caption{\label{tab:results}Stemma reconstruction by exact MWIS and by existing methods, on an artificial tradition of known history and on a real one. Heinrichi: skeleton score of the reconstructed tree against the true tree (\%, higher is better). Donatus: number of the editor's 12 families recovered exactly.}
\begin{ruledtabular}
\begin{tabular}{lcc}
\toprule
Method & Heinrichi & Donatus \\
\midrule
Neighbour joining & 67.9 & 8 \\
Neighbor-Net, greedy tree$^{b}$ & 71.8 & 8 \\
Bayesian (MrBayes), consensus & 82.2$^{c}$ & 8 \\
Camps--Cafiero & --$^{d}$ & 5 \\
Roelli--Bachmann \emph{Leitfehler} & 68.6 & 9 \\
RHM, on the words$^{e}$ & 77.8--83.1 & -- \\
Semstem & 81.1--83.9 & 5--6 \\
Maximum parsimony & 81.7 & 9 \\
MWIS, pairwise heuristic$^{a}$ & 82.2 & 8 \\
MWIS, exact joint completion & 79.1--82.3$^{f}$ & 8--10$^{f}$ \\
Hybrid, $\theta=0.8$ / $0.9$ & 81.7 / 81.7 & 10 / 10 \\
\bottomrule
\end{tabular}
\end{ruledtabular}
{\raggedright\footnotesize $^{a}$Pairwise relaxation, nearest-neighbour extension, greedy filter. $^{b}$Splits of the network kept in decreasing weight if compatible; on Donatus 10 of the 12 families are among them, 8 in the tree. $^{c}$Not converged on Heinrichi; the consensus is the same at $10^6$ and $4\times10^6$ generations, so the score is indicative only. $^{d}$Returns no stemma in its automatic mode on Heinrichi. $^{e}$Reads the aligned words; the other methods read coded variant states, so this is not an identical-input comparison. Complete aligned witness transcriptions were unavailable for this study. $^{f}$Range over the optimal solutions, all equally supported: 12 on Heinrichi, enumerated in full, and at least 200 on Donatus, where the enumeration stopped at 200, so the Donatus range is over the 200 examined; no single one is reported (Sec.~\ref{sec:conclusion}). RHM: three restarts; Semstem: three seeds; their ranges are local optima of the search. The implementation and settings of every baseline are in the code and data deposit (see Data availability).\par}
\end{table}
%=====================================================================

Two traditions serve as benchmarks. \emph{Heinrichi} is an artificial tradition: a Finnish text copied by hand by volunteers under controlled conditions, released for the 2007 computer-assisted stemmatology challenge with its true copying history (37 witnesses, 1{,}208 units;~\citealp{RoosHeikkila2009}). We score trees with a re-implementation of the challenge's skeleton criterion, which reproduces exactly the value published on the challenge site for the organisers' own neighbour-joining file, 67.887~\citep{CASC2007}; the summary values of~\citet{RoosHeikkila2009} are on a different scale, so every method is rerun here rather than quoted. \emph{Donatus} is the \emph{Passio} of Donatus of Arezzo (BHL 2294), edited by Pierluigi Licciardello from 54 manuscripts with a full negative apparatus~\citep{Licciardello2018}; the apparatus was transcribed twice independently from the page images by AI agents, disagreements being decided on the page images (cell agreement 99.6\%, lemma cells included), into a 54$\times$1{,}253 matrix, which we compare with the editor's stemma, which has 12 named families. The apparatus cites groups of manuscripts by family sigla, which we expand to their members; the matrix therefore carries part of his stemma, and recovery of his families measures consistency with his own evidence, not independent confirmation. \emph{Lawrence} is the prologue \emph{In divinis laudibus}, part of which the bifolium carries (Fig.~\ref{fig:plates}b); its seven witnesses, collated by us, are few enough to inspect by hand. The baselines are neighbour joining, the Neighbor-Net split network~\citep{BryantMoulton2004} read as a greedy tree, Bayesian inference with MrBayes~\citep{Ronquist2012}, maximum parsimony, the method of Camps and Cafiero~\citep{CampsCafiero2015}, the \emph{Leitfehler} weighting of Roelli and Bachmann~\citep{RoelliBachmann2010}, and two methods designed for stemmatology, the compression-based method of Roos, Heikkilä and Myllymäki (RHM)~\citep{RoosHeikkilaMyllymaki2006} and Semstem~\citep{RoosZou2011}. We also test a hybrid of exact compatibility and parsimony, built on the bootstrap~\citep{Felsenstein1985}. The bootstrap draws the units of the collation at random, with replacement, to form new collations of the same size, computes the best-supported stemma of each, and gives every family a \emph{stability}, the share of these replicates in which it appears; a family that rests on a few units has low stability. The hybrid keeps the families whose stability is at least $\theta$ and lets parsimony resolve the rest; $\theta$ is chosen on Heinrichi (0.8--0.9), which is therefore a development set for the hybrid, and applied unchanged to Donatus.

Table~\ref{tab:results} summarises. On the benchmark with known truth, the optimal exact-compatibility solutions (79.1--82.3) overlap parsimony (81.7) and the Bayesian consensus (82.2); neighbour joining, the split network and Roelli and Bachmann's \emph{Leitfehler} weighting, the method closest to ours in principle, fall well below. Semstem and RHM, designed for stemmatology, do as well on the artificial tradition, but Semstem recovers only 5 or 6 of the editor's 12 families on Donatus, against 8 to 10 for the 200 exact optima examined, and RHM cannot be run on a coded apparatus. The hybrid matches parsimony on Heinrichi (81.7) and recovers 10 of the 12 families on Donatus. With the family sigla of the Donatus apparatus left unexpanded, no method recovers more than two of the twelve. Exact compatibility and parsimony make largely the same errors on Heinrichi, because they read the same evidence and inherit the same ambiguities. The correction for fragmentary witnesses (Sec.~\ref{sec:partial}) is small: on Heinrichi the pairwise relaxation exceeds the exact optimum by one unit of weight (537 against 536, two genuine triple obstructions), and on Donatus it is already exact.

On Donatus, the bootstrap (200 replicates) separates a robust lower stemma (six of the editor's families recovered in every replicate or nearly) from a top division that the data do not decide: the editor's top family $\nu$ appears in fewer than 10\% of replicates, the closest stable split being $\mu$ against the other 41 witnesses (Jaccard similarity 0.71 between that 41-witness side and $\nu$, stability 0.91), and the families of the hybrid trees that cross the editor's families almost all cross $\nu$. The best tree that contains all twelve of the editor's families weighs 569 units against 587 for the optimum: imposing them costs 18 units of support, 15 of them for $\nu$. No removal of one to three witnesses resolves those crossings, so the disagreement is diffuse rather than due to a single contaminated manuscript. The bifolium L is not among the edition's witnesses; collated by hand against the apparatus, its text of the passion sits inside $\upsilon$ with the Trento passional (T2 in the edition, T in Fig.~\ref{fig:plates}c), with which it also shares seven innovations in the prologue (Sec.~\ref{sec:lawrence}). 

The procedure also names what to check: the rejected families and the manuscripts whose removal would reconcile them (a \emph{conflict report}), and the loci behind each disputed family. We checked three of the implicated witnesses against digitised images, at about twenty loci (Vatican City, Biblioteca Apostolica Vaticana, Vat.\ lat.\ 1271; Paris, Bibliothèque nationale de France, lat.\ 5296 and lat.\ 5296 B): the edition's apparatus is accurate at every locus read, and a word-order inversion shared by three witnesses that the editor places in different branches is confirmed in the two of them read on images. Lawrence is discussed in Sec.~\ref{sec:lawrence}.

%=====================================================================
\subsection{Reading a stemma with the evidence map}
\label{sec:evidencemap}
%=====================================================================

A published stemma does not say which of its branches the editor would defend firmly and which are conjecture. The quantities computed above give each family of an existing stemma three numbers that an editor can read directly:

\begin{itemize}
\item \emph{support}: the number of variation units at which the family's legible members share a reading against every other legible witness. These are shared \emph{readings}, not established errors: whether a reading is an innovation that binds, an inherited reading, or a change easily made twice is the editor's judgement, which the count does not replace;
\item \emph{crossing weight}: the summed support of the candidate families that cross it, so that a unit proposing several crossing families counts once for each; these families form a \emph{conflict report}, a list of candidates for examination; whether a conflict reflects coincidence, contamination or a misplaced family is decided by philological examination, not by the count;
\item \emph{stability}: the share of bootstrap replicates (Sec.~\ref{sec:experiments}) whose exact optimum contains the family; the solver certified the optimum of every replicate, and where several optima tie, the solver's deterministic choice is counted. Stability measures sensitivity to resampling under the model; it does not validate the model or the family.
\end{itemize}

\begin{table}[tbp]
\caption{\label{tab:evidence}Evidence for each family of the editor's stemma of the \emph{Passio Donati} (BHL 2294; 54 witnesses, 1{,}253 units). Size: number of witnesses. Support: units at which the family's members share a reading against all other witnesses. Crossing: summed support of the families that cross it. Stability: share of 200 bootstrap replicates whose optimum contains the family. Reading: class of the family, from support and stability.}
\begin{ruledtabular}
\setlength{\tabcolsep}{2.5pt}
\begin{tabular}{lrrrrl}
\toprule
Family & Size & Support & Crossing & Stability & Reading \\
\midrule
$\lambda$ & 2 & 188 & 89 & 1.00 & solid \\
$\kappa$ & 4 & 44 & 120 & 1.00 & solid \\
$\phi$ & 5 & 26 & 124 & 1.00 & solid \\
$\rho$ & 3 & 24 & 269 & 0.93 & solid \\
$o$ & 3 & 18 & 250 & 1.00 & solid \\
$\sigma$ & 3 & 16 & 132 & 0.99 & solid \\
$\mu$ & 13 & 8 & 502 & 0.91 & few but stable \\
$\xi$ & 9 & 7 & 301 & 0.92 & few but stable \\
$\pi$ & 2 & 14 & 149 & 0.54 & contested \\
$\tau$ & 7 & 7 & 439 & 0.41 & weak \\
$\upsilon$ & 4 & 3 & 176 & 0.18 & weak \\
$\nu$ & 29 & 0 & 805 & $<$0.1 & unattested \\
\bottomrule
\end{tabular}
\end{ruledtabular}
\end{table}

Table~\ref{tab:evidence} applies this to the twelve families of Licciardello's stemma of the \emph{Passio} of Donatus~\citep{Licciardello2018}. It sorts them into five classes that an editor can act on, by the following rule: \emph{solid}, stability at least 0.90 and support at least 15; \emph{few but stable}, stability at least 0.90 and support below 15; \emph{contested}, stability from 0.5 to below 0.90; \emph{weak}, stability below 0.5; \emph{unattested}, no support. Two cautions apply. Support falls with the size of a family, since a single member that departs from the shared reading, or was corrected from another exemplar, removes the unit; and crossing weight is a sum over rival families that may conflict among themselves, so it measures how many rivals there are rather than how strong they are.

\emph{Solid families} ($\lambda,\kappa,\phi,\rho,o,\sigma$) are attested by many units and survive nearly every resampling; on this evidence they are not in question. \emph{Few but stable} families ($\mu,\xi$) rest on a handful of readings that nothing strongly contradicts, so each stands or falls with those readings, which an edition can cite. In \emph{contested or weak} families ($\pi,\tau,\upsilon$) the evidence is divided, and the method points to what would decide it. For $\upsilon$ (F20, L5, Mi, T2), the same witnesses without L5 form a family present in 99\% of replicates, and that smaller family joins $\sigma$ (Ad, L3, R4) in 77\% of replicates, a grouping that the published stemma does not contain. The editor's question is then precise: where does L5 belong, and are $\upsilon$ and $\sigma$ one branch? It can be settled by re-reading these seven witnesses at the loci listed by the method. Finally, the top family $\nu$ has no supporting unit: no reading is shared by its 29 members against all the other witnesses, while 805 units propose families that cross it. For a family of 29 this is weak evidence against it, by the first caution above, and the editor may have inferred $\nu$ from many partial agreements or from criteria other than readings; what the map shows is that $\nu$ is inferred, not observed in the apparatus.

Each of these flags leads back to readings that can be checked in the manuscripts. For $\upsilon$, the edition's own apparatus records the exception: at six loci it gives the reading of $\upsilon$ ``tranne L5'', without L5 (\emph{deberi sibi}, XIV.2), against three readings shared by $\upsilon$ with L5. The editor noted the exception locus by locus; the map gathers it into one question, where L5 belongs. Across the top division, the inversion \emph{quem perquirens studiose} (VII.4) is shared by P1, R2 and V2, which the edition places in different branches, and is confirmed on the images of two of them (Sec.~\ref{sec:experiments}). Such an agreement proves no common exemplar, but it is the kind of locus the map lists for a second reading.

The method therefore has two uses. Where no stemma exists, it proposes one, the best-supported tree, as for the prologue in Sec.~\ref{sec:lawrence}. Where an editor's stemma exists, it gives each branch its evidence, as here for Donatus. Support and crossing are counted on the collation and apply to the editor's families as drawn; only stability depends on the method's own optimum.

%=====================================================================
\subsection{The prologue \emph{In divinis laudibus} and its seven copies}
\label{sec:lawrence}
%=====================================================================

Donatus tests the method against a published stemma. The prologue \emph{In divinis laudibus}, part of which the bifolium L carries (Fig.~\ref{fig:plates}b), shows the other use: building a stemma where none existed, from a full collation of the copies, which the method does not replace. The prologue (BHL 4759) stands before the passion of St Lawrence in Italian passionals of the eleventh and twelfth centuries. Seven copies survive (Appendix~\ref{app:witnesses}), among them L, whose text is close to that of the passional now in Trento (Fig.~\ref{fig:plates}c): T and L share seven innovations in the prologue, among them the word order \emph{puniri capite} (p.~125b, line 25), where the other five witnesses read \emph{capite puniri}. The matrix used here is generated from our collation (negative apparatus, first hands, spelling variants counted with the lemma) and has 75 informative units.

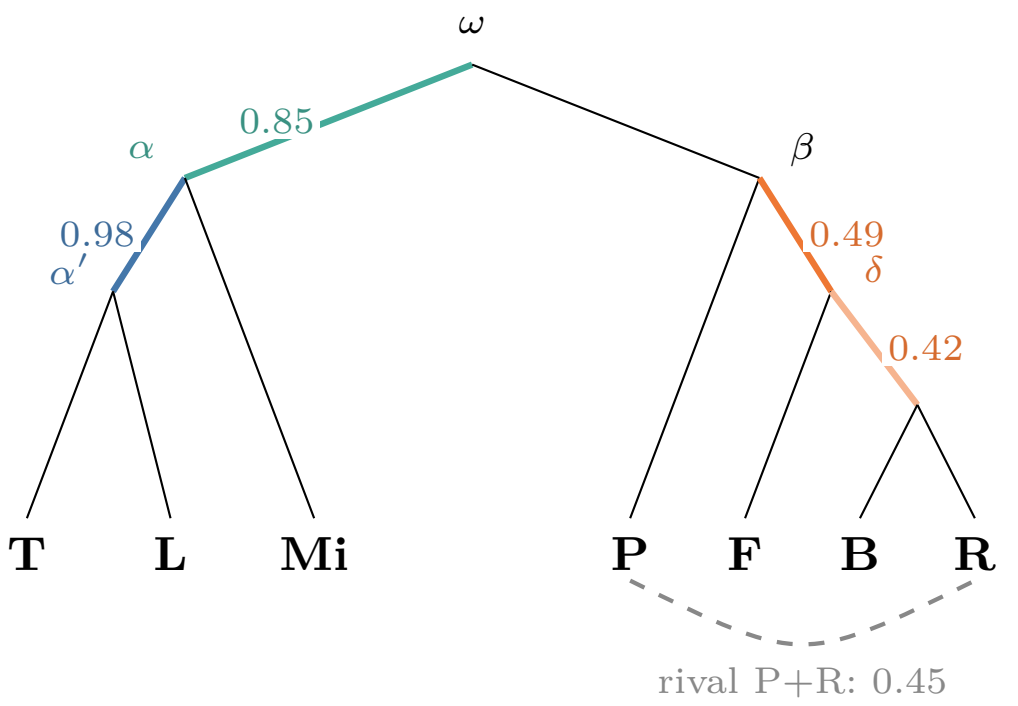
\begin{figure}[tbp]
\centering
\begin{tikzpicture}[x=0.95cm,y=0.75cm,every node/.style={font=\small},
  w/.style={font=\small\bfseries}, s/.style={font=\scriptsize,fill=white,inner sep=1pt}]
  \coordinate (om) at (3.5,4);
  \coordinate (al) at (1.5,3); \coordinate (alp) at (1,2);
  \coordinate (be) at (5.5,3); \coordinate (de) at (6,2); \coordinate (br) at (6.6,1);
  \draw[colABC,very thick] (om)--(al) node[s,midway,left=1pt,text=colABC!85!black]{0.85}; \draw (om)--(be);
  \draw[colAB,very thick] (al)--(alp) node[s,midway,left=1pt,text=colAB!90!black]{0.98};
  \draw (al)--(2.4,0) node[w,below]{Mi};
  \draw (alp)--(0.4,0) node[w,below]{T}; \draw (alp)--(1.4,0) node[w,below]{L};
  \draw (be)--(4.6,0) node[w,below]{P};
  \draw[colDE,very thick] (be)--(de) node[s,midway,right=1pt,text=colDE!90!black]{0.49};
  \draw (de)--(5.4,0) node[w,below]{F};
  \draw[colDE!55,very thick] (de)--(br) node[s,midway,right=1pt,text=colDE!90!black]{0.42};
  \draw (br)--(6.2,0) node[w,below]{B}; \draw (br)--(7.0,0) node[w,below]{R};
  \draw[dashed,colRej,thick] (4.6,-0.55) .. controls (5.8,-1.3) .. (7.0,-0.55);
  \node[font=\scriptsize,text=colRej] at (5.8,-1.45) {rival P+R: 0.45};
  \node[font=\scriptsize] at (3.5,4.35) {$\omega$};
  \node[font=\scriptsize,text=colABC!85!black] at (1.2,3.25) {$\alpha$}; \node[font=\scriptsize,text=colAB!90!black] at (0.7,2.2) {$\alpha'$};
  \node[font=\scriptsize] at (5.8,3.25) {$\beta$}; \node[font=\scriptsize,text=colDE!90!black] at (6.3,2.2) {$\delta$};
\end{tikzpicture}
\caption{\label{fig:lawrence}Stemma of the prologue \emph{In divinis laudibus} (BHL 4759) from exact MWIS on 75 units, drawn with the root of the stemma drawn by hand. Numbers: bootstrap stabilities ($B=200$); dashed: the rival pair P+R. T: Trento; L: the bifolium; Mi: Milan; F: Florence; P and B: Lucca; R: Rome.}
\end{figure}

Before the method was applied, we drew a stemma by hand from shared innovations. The exact optimum (Fig.~\ref{fig:lawrence}) has its groups: the pair T--L ($\alpha'$), its association with Mi, and $\delta$ (F, B, R). The two differ inside $\delta$. The optimum joins B and R, but only in 42\% of the bootstrap replicates; the pairing P+R, which crosses the boundary of $\delta$, is chosen in 45\%: the data do not decide between them. Removing P or R from the data removes the conflict. The stemma drawn by hand reached the same point by reading: R's run of agreements with P is taken there as correction from a second exemplar, $\delta$ is drawn with its lower node unresolved, and R's double affiliation is stated. The bootstrap gives the weight of each alternative.

%=====================================================================
\subsection{Performance and role of generative language models}
\label{sec:llm}
%=====================================================================

Large language models can read a collation and propose a stemma, which raises an obvious question: why build exact methods at all? We gave fifteen current models, closed and open-weight, the same matrix as the exact method, with every siglum replaced by an anonymous label, asked for the families in a fixed format, and scored the trees with the same criteria; the prompt, the parsing of answers and the scoring were fixed in advance and identical for all models; the protocol and the full results are in Appendix~\ref{app:llm}, and every answer in the code and data deposit (see Data availability). One model, Claude Fable 5.2, scores 83.3 on Heinrichi over the three runs in the table (82.0 to 84.2) and 82.5 over all six runs made (77.3 to 84.3), at the top of the range of the exact optima, and recovers 8 or 9 of the editor's 12 families on Donatus, within the range of the 200 exact optima examined (8 to 10); all the others that returned a tree are 7 to 30 points below the exact methods, and more reasoning did not give better trees.

A good tree, however, is not all that an edition needs. A stemma decides which readings enter the text, so each of its branches must be defended by readings that a reader can check, and the answer must be the same when someone else runs it. The models' reasons do not meet that standard. Of 306 pieces of evidence that eight models cite for their families, checked by program against the matrix they were given, 63\% hold for every witness named and 30\% for only part of it, and only 41\% meet the method's own definition of support, shared by no witness outside the family. The model with the best trees cites mostly undiagnostic evidence, the model whose citations are most reliable builds one of the weakest trees, and scores vary from run to run. An editor could use none of these answers without re-deriving every claim. The exact method returns an optimal stemma with a certificate that no compatible set of families has more support, and with the units behind each family and the families it defeats; support and crossing can then be computed for any tree, whatever method drew it. The two can be combined: a model proposes families, and the optimiser checks which are jointly consistent and what supports them.

%=====================================================================
\section{Critical editions revisited}
\label{sec:founding}
%=====================================================================

With the method tested, this section uses it to examine the editors' work. Read back from their own apparatus, the stemmata of a whole series of editions show which of their branches rest on the variants and which on the editor's judgement, how far later editions have changed them, and whether the two-branched tops that Bédier found suspicious are in the evidence. The material is the \emph{Monumenta Germaniae Historica} (MGH): the \emph{Auctores antiquissimi} (1877--1919), fifteen volumes of late antique and early medieval Latin authors, and the \emph{Scriptores rerum Merovingicarum} (1884--1920). Their editions set the standard of the genealogical method in its classical period, their negative, regular apparatus can be read by machine with every unit traceable to its line, and their page images are freely available. Here, unlike Sec.~\ref{sec:results}, families are counted as shared departures from the edited text (the rooted variant of Sec.~\ref{sec:complete}). The extracted apparatus, the code, the shelfmarks and the details of every analysis in this section are in the code and data deposit (see Data availability).

The lower families of a stemma are the groups of closely related manuscripts within a branch; its top division is the split of all the witnesses into the branches that descend directly from the archetype. Sec.~\ref{sec:mommsen} reads one edition in full, Mommsen's \emph{Getica}, chosen because a later edition gives an independent check on what the map finds. Sec.~\ref{sec:mgh} repeats the reading on four more editions, to see whether the result depends on the editor. Both concern the families an editor uses to establish his text. Bédier's objection concerns the top division, and Sec.~\ref{sec:bedier} tests it across every text of the series whose editor gives an explicit one. Sec.~\ref{sec:lai} takes Bédier's own text, where the evidence is unusually clean, to show what the readings can and cannot decide about a top; Sec.~\ref{sec:thegan} shows, on an edition whose editor states his reasons, where the remaining decision lies.

\begin{table}[!b]
\caption{\label{tab:mommsen}Families of manuscripts found in Mommsen's apparatus of Jordanes' \emph{Getica}, compared with his stemma. Each row is a family of the exact optimum. Departures: places where its members share a departure from Mommsen's text. Stability: share of 200 bootstrap replicates whose optimum contains the family. Mommsen: the order of his stemma that the family matches.}
\begin{ruledtabular}
\begin{tabular}{lrrl}
\toprule
Family & Departures & Stability & Mommsen \\
\midrule
O B & 436 & 1.00 & order II \\
X Y Z & 295 & 1.00 & order III \\
X Y & 157 & 1.00 & --- \\
H P V & 37 & 1.00 & --- \\
S O B & 36 & 0.97 & order II \\
H P V L & 34 & 0.97 & order I \\
A X Y Z & 34 & 0.90 & A in order I \\
H P & 27 & 0.78 & --- \\
\bottomrule
\end{tabular}
\end{ruledtabular}
\end{table}

\subsection{Mommsen's \emph{Getica}, checked against a later edition}
\label{sec:mommsen}

The \emph{Getica} of Jordanes is the test case because its stemma has been revised since: the edition of the historian Theodor Mommsen~\citep{Mommsen1882} was followed a century later by Grillone's, built on a Palermo manuscript Mommsen did not know~\citep{GiuntaGrillone1991,Grillone2017}. Where the map departs from Mommsen, the later edition says whether the departure is noise or a point a better-informed editor also changed. Mommsen reports eleven witnesses (Table~\ref{tab:editions}), among them A, the Ambrosianus; his stemma has three orders, I = HPVL with A, II = SOB, III = XYZ, the first two sharing an archetype against the third. A parser written for this apparatus yields 4{,}611 units. So as not to carry Mommsen's stemma into the result, departures were also counted from the majority reading instead of his lemma; the two references coincide in 93\% of units and give the same families.

Mommsen's three orders come out of the apparatus alone, at stability 0.97--1.00 (Table~\ref{tab:mommsen}). The map departs from him at two points. The first is A, which Mommsen derives from P or a twin of P, ascribing its agreements with XYZ to the scribe's urge to emend. Where A agrees with Mommsen's text it agrees with order I, but where it departs it departs with the other orders: its strongest rooted family is A+XYZ (0.90), and among substantive departures it shares 22 with order I alone, 22 with order II alone, 34 with order III alone and 37 with II and III together, the profile of mixed affiliation. The readings Mommsen reports for AXYZ stand in the first hand of the Ambrosianus at the three decisive places where we read them, a spot check of the family rather than a verification; a fourth place confirms A with P, and a fifth is an unresolved abbreviation. The second is the top: his apparatus holds almost no evidence for a common ancestor of orders I and II (2 units in which one reading is shared by every cited member of orders I and II and by none of III, among the 4{,}587 whose edited reading is identified, against 167 for II and III; those 167 are spread over 18 different sets of witnesses, because S, Z and A are cited only intermittently, so no single family of Table~\ref{tab:mommsen} collects them). Grillone's edition, read only after these results, changes the stemma at the same two points: it drops the common ancestor of I and II, and gives A a branch of its own with contamination, close to, though not the same claim as, the mixed affiliation found here. On this edition, then, the map recovers the editor's families and flags the two places where a later editor revised him.

\subsection{Four more editions of the \emph{Monumenta}}
\label{sec:mgh}

\begin{table*}[b]
\caption{\label{tab:mgh}What the evidence map finds in five editions of the \emph{Monumenta Germaniae Historica}, read back from their own apparatus. Wit.: witnesses cited by siglum. Dep.: shared departures from the edited text. Editor's groups recovered: families the editor states, with their bootstrap stability (share of 200 replicates). Raised by the map: where the map departs from the editor. All rows come from one extractor written for the five editions; for the Getica its figures differ slightly from those of the dedicated parser.}
\begin{ruledtabular}
\footnotesize
\begin{tabular}{p{2.6cm}p{2.0cm}rrp{4.6cm}p{5.0cm}}
\toprule
Text & Editor & Wit. & Dep. & Editor's groups recovered (stability) & Raised by the map \\
\midrule
Jordanes, \emph{Getica} & Mommsen 1882 & 11 & 2{,}414 & OB, XYZ 1.00; SOB 0.99; HPVL 0.98 & A with XYZ (0.94): mixed affiliation (Sec.~\ref{sec:mommsen}) \\
Eutropius & Droysen 1879 & 7 & 1{,}722 & LO, PD, FG 1.00 & LO+PD (81 departures, 1.00): the agreements Droysen and Mommsen dispute \\
Sidonius, letters & Luetjohann, Leo 1887 & 6 & 2{,}149 & MTCFP, CF 1.00 & Leo's pair MT absent: M sides with CFP in 308 departures, mostly spellings, all checked ones in M's first hand \\
Cassiodorus, \emph{Var.} I--IV & Mommsen 1894 & 13 & 1{,}032 & EF, XEF, PD, LR 1.00; LRK 0.88 & class II not one family; class I core weak (MN 0.59); X with EF, as Mommsen says \\
\quad \emph{Var.} V--VII & & 11 & 700 & EF, TU, LR, KEF 1.00 & PA 0.70 (A a copy or twin of P) \\
\quad \emph{Var.} VIII--XII & & 12 & 1{,}696 & EF, GHI, GH, ZN 1.00; KEF 0.67 & B, the leading witness of class IV, joins no family \\
Fortunatus, \emph{Carmina} & Leo 1881 & 11 & 2{,}648 & BL 1.00, RF 0.98, DG 0.94 & ACM 1.00: 55 places printed against the best manuscripts \\
\bottomrule
\end{tabular}
\end{ruledtabular}
\end{table*}

One edition could be a lucky case. We therefore read four more editions of the \emph{Auctores antiquissimi}, by editors whose prefaces state how the manuscripts group: Droysen's Eutropius~\citep{Droysen1879}, Leo's Fortunatus~\citep{Leo1881}, Sidonius, edited by Luetjohann and completed after his death by Leo, who drew its stemma~\citep{Luetjohann1887}, and Mommsen's \emph{Variae} of Cassiodorus~\citep{Mommsen1894}. Here a single extractor, run on the optical character recognition (OCR) of the page images, keeps the groups that share a departure from the edited text, and the Getica was re-read with it for comparison. On twelve printed pages transcribed from the images by an AI agent, not checked by a human reader, it recovers 84\% of the 200 such groups, with precision of at least 90\%; these are development figures, since the pages guided the repairs.

The pattern is the same in every edition (Table~\ref{tab:mgh}). The pairs and classes on which each editor built his text come out, most at stability 0.95 or above: Droysen's three families, Leo's three interpolated pairs in Fortunatus, Mommsen's classes in each part of the \emph{Variae}, including relations he states in passing, such as X sharing the errors of EF and the later recension EF following a codex like K. Where the map departs from an editor, it points to a question he had raised himself. In Eutropius, the 81 places where Droysen's classes B (LO) and C (PD) agree against A are exactly those over which Droysen (contamination of C from B) and Mommsen, in a note to Droysen's own preface, disagree. In Sidonius, M sides with the interpolated branch rather than with T. The manuscript (Florence, Laurenziana, S. Marco 554) confirms these readings in its first hand, but they are mostly spellings and one haplography, errors that copyists make independently; Leo had grouped M with T by their shared gaps and the order of the letters, evidence that an apparatus of readings does not record. In this case the map shows which kind of evidence the stemma rests on. In Fortunatus, 55 places where Leo's text departs from his best manuscripts A, C and M are worth listing for any revision. In none of these editions does the map overturn the editor. It recovers the groups on which he built his text, and where it departs from him it lands on a question he had raised himself, with the list of readings that would decide it; for anyone revising one of these editions, that list is where to start. These five editions were chosen because their stemmata looked testable, and what they test is the lower families. Whether editors' top divisions are equally well founded needs a test that takes every text of the series, not a selection.

\subsection{Are editors' top divisions in their evidence?}
\label{sec:bedier}

\begin{table}[!b]
\caption{\label{tab:bedier}Whether the editors' own apparatus supports their top divisions, in the fourteen texts of the series test, all edited in the \emph{Auctores antiquissimi} of the \emph{Monumenta Germaniae Historica} (1877--1919). Supported: every branch of the top division is a stable family of the apparatus. Contradicted: a stable family crosses a branch. Silent: neither.}
\begin{ruledtabular}
\begin{tabular}{lp{0.62\columnwidth}}
\toprule
Verdict & Texts \\
\midrule
Supported (4) & Eugippius, \emph{Vita Severini}\newline Marcellinus Comes, \emph{Chronicon}\newline Sidonius, \emph{Carmina}\newline Victor of Vita, \emph{Historia persecutionis} \\
\addlinespace
Contradicted (4) & Ennodius, works\newline Fortunatus, \emph{Vita Martini}\newline Paul the Deacon, \emph{Historia Romana}\newline Prosper, \emph{Epitoma chronicon} \\
\addlinespace
Silent (6) & Claudian, \emph{Carmina maiora}\newline Claudian, \emph{De raptu Proserpinae}\newline \emph{Historia Brittonum}\newline Isidore, \emph{Chronica maiora}\newline Isidore, \emph{Historia Gothorum}\newline Jordanes, \emph{Romana} \\
\bottomrule
\end{tabular}
\end{ruledtabular}
\end{table}

Bédier's objection concerns the top of the stemma. Of 110 stemmata he collected from editions of French texts, 105 divide in two at the top, which he found incredible: ``Un arbre bifide n'a rien d'étrange, mais un bosquet d'arbres bifides, un bois, une forêt? \emph{Silva portentosa}''~\citep[171--172]{Bedier1928}. Is the division in the evidence, or in the method? The evidence map lets the question be put to each edition: does the editor's own apparatus support the top division he draws?

To answer it for a whole series rather than a selection, we recorded the stemma of every text of the \emph{Auctores antiquissimi}, 71 texts in fifteen volumes, from the editors' prefaces and printed stemmata. The record, released in the code and data deposit, shows how the editors of a founding series drew the genealogy of their texts, and it is the base of the counts below; 46 texts have a top division. The 4{,}091 pages of the texts that could be located in the page images were transcribed by a vision model and their apparatus extracted (see Data availability). A text enters the test only if its apparatus holds enough evidence to decide, at least four regularly cited witnesses and 200 units with a shared departure, and if its extraction passes a check on a page drawn at random; these criteria were fixed before any text was extracted. Fourteen texts pass and have an explicit top (Appendix~\ref{app:series}). The five editions of Secs.~\ref{sec:mommsen}--\ref{sec:mgh} are not among the fourteen texts, although four of these come from the same volumes, and Claudian and Isidore count two texts each, so the fourteen are not fully independent. Each branch of an editor's top division is \emph{supported} if it is a stable family of the optimum (stability at least 0.90), \emph{contradicted} if a stable family of at least five units crosses it, and \emph{silent} otherwise; a division is contradicted if any of its branches is, and supported if all are (branches of a single witness are not classified).

The answer is mixed: four divisions are supported, four contradicted and six silent (Table~\ref{tab:bedier}). Victor's top is supported, his branch BLV being a stable family of 149 units. Vogel's in Ennodius is contradicted: B, alone on one side, shares 211 departures with V and L on the other. Claudian's are silent: the lower families are stable, and nothing in the apparatus groups the top. Silence is the commonest outcome, and it is Bédier's suspicion in a weak form: the top is not disproved, but the apparatus does not establish it. The author read 60 extracted departures on the page images, 40 of them behind the contradictions, blind among the others, and all 60 are recorded with the right manuscripts; under simulated extraction errors twelve of the fourteen verdicts are unchanged; and the pattern is the same without Prosper, admitted as a stated exception. A supported top is compatible and stable, not shown to be free of the over-resolution discussed in Sec.~\ref{sec:conclusion}. Where a top is contradicted, the method does not say whether the crossing agreement is contamination or a kinship the editor missed, but it lists the readings.

A top the apparatus does not prove is not therefore arbitrary. When the same witnesses are divided at random into branches of the same sizes, most of these divisions are contradicted, against four of the editors' thirteen with two stated branches ($p\approx10^{-5}$). The advantage comes from the lower families. Random divisions that keep intact the smallest stable families, those lying inside another, still do worse than the editors' ($p=0.008$); random divisions that keep every stable family intact cannot be contradicted, and among them the editor's is no better supported than the others, and in three texts worse. Editors built their tops on solid lower families; the apparatus rarely shows why they chose one top among those the families allow. All four supported tops set a single manuscript against the rest, so only one branch had to earn its support.

Bédier's count also holds in this series. Over the \emph{Auctores antiquissimi} and the \emph{Scriptores rerum Merovingicarum} (239 texts), 90 of 94 explicit tops are two-way. The model of copying and loss of the Hebraist Michael Weitzman predicts about 71\% for Latin and 77\% for Greek traditions~\citep[303]{Weitzman1987}, and a birth--death model calibrated on Old French romances 57--67\%~\citep{Camps2025}; the observed 96\% exceeds both, so loss, as these models represent it, does not seem to account for all of it~\citep{GuidiTrovato2004,Trovato2017}. Three editors, Krusch, Levison and Mommsen, drew 70 of the 94, so the count describes these editions rather than testing either model. Two exploratory checks fit Bédier's own explanation, that editors first sort manuscripts into three families or more and then prune the scheme to two~\citep[172--173]{Bedier1928}: the evidence at the top is thin and evenly balanced, and the edited text follows one preferred branch.

Later editions settle nothing either way. Nine of the fourteen texts have a later top to compare. Of the three contradicted texts that have a later edition, two were revised, the \emph{Vita Martini}~\citep{Kay2020} and Paul~\citep{Crivellucci1914}, and Prosper was kept~\citep{BeckerKoetter2016}. Of the six others, Claudian's later editor gives up a stemma for both texts, as an open, contaminated tradition~\citep[x--xi]{Hall1985}; a later study regroups Sidonius' poems~\citep{Dolveck2020}; and the later editions of Eugippius~\citep{Knoell1886}, Victor~\citep{Petschenig1881} and Isidore's Gothic history~\citep{RodriguezAlonso1975} keep their tops. Outside the test, the Getica, contradicted by the same rule, was revised~\citep{Grillone2017}. These cases are too few for a test.

\subsection{Bédier's own case: the \emph{Lai de l'Ombre}}
\label{sec:lai}

The series test depends on extraction from printed pages and on texts whose evidence is often thin. Bédier's own text allows the question to be asked once under the best conditions: a short poem, a clean apparatus, three published classifications that disagree. If the readings cannot fix the top there, the limit is not one of data quality.

Bédier's objection grew out of Jean Renart's \emph{Lai de l'Ombre}, preserved in seven manuscripts, A to G. In his first edition~\citep{Bedier1890} Bédier divided them into two families, ABCG and DEF. Gaston Paris, his teacher and a founder of Romance philology in France, replied that the evidence for DEF did not hold and proposed three families, ABCG, DF and E~\citep{Paris1890}. Quentin, applying his own method, found three branches, AB, CG and DE, with F derived from the last two, and objected to Bédier's grouping the ``accord constant de CGDE contre AB''~\citep[160--162]{Quentin1926}. We read the variant notes of Bédier's second edition~\citep{Bedier1913} from the page images, as for the \emph{Monumenta}, and obtained 806 notes covering the whole poem, split into 843 units, one for each word or phrase a note treats, so that no unit puts a witness in two groups; on a page drawn at random and transcribed from the image, the extraction recovers 42 of its 43 readings, every shared group among them, with nothing added.

The readings fix the unrooted tree ((A,B),(C,G),((D,E),F)). Its four families are heavy and stable: CG share 176 departures, DE 41, AB 38 and DEF 24, all at stability 0.98 or above, and every competing grouping is lighter and crossed by one of them (EF, 23, and DF, 13, cross DE). The best tree that differs from it replaces DEF by CDEG (10) and scores 265 against 279, so the optimum is unique on this apparatus by 14 shared departures. That is a property of the collation, not a statistical guarantee that the tree is the true history.

Of the three published classifications, one conflicts with this tree. Paris's joins D with F, and DF (13 departures) crosses DE (41). Nor is DE likely to be a chance agreement of two of three copies made independently from one exemplar, the case discussed in Sec.~\ref{sec:conclusion}: under equal rates of shared error, a lead as large as DE's over EF arises with probability 0.004, with unequal rates, read from each witness's singular departures, a three-way branching is still rejected unless coincident errors are almost absent (with none at all, $p=0.08$)~\citep{JurczakCoincidence}.

The other two do not conflict with it. Bédier's two families, Quentin's three branches AB, CG and DE with F beside DE, and a division of AB against the rest are all this same unrooted tree, rooted next to DEF, at the centre, or next to AB; Quentin also derived F from two branches, a second exemplar that no tree can show, and the comparison concerns only the tree. Which root is best depends on which readings count as original. Against Bédier's printed text the root next to AB wins, with 276 shared departures against 258 for Bédier's root and 255 for the centre, because Bédier printed A's text: wherever A and B keep it, the agreement of the other five counts as a shared departure, Quentin's ``accord constant de CGDE contre AB''. Against the majority reading the three rootings tie.

Shared readings decide the groups; the root follows from which readings are taken as original, a judgement the readings alone do not make. In Bédier's own case, then, the question he raised is the question of the root, and the readings leave it open.

\subsection{Thegan's \emph{Gesta Hludowici}: where the editor decides the root}
\label{sec:thegan}

If the readings leave the root open, an editor must close it, and the evidence map can show where. Thegan's \emph{Gesta Hludowici}, a short life of Louis the Pious written by a chorbishop of Trier in the emperor's lifetime, is a case where the editor states the judgement that does so. Tremp's edition~\citep{Tremp1995} cites twelve witnesses, among them Pithou's first edition of 1588 (Pi), printed from a manuscript now lost; seven carry the whole text. From the apparatus these seven give one unrooted tree, ((W,(T,P)),((B3,Pi),(L2,B4))), every split at stability 1.00, whether departures are counted against Tremp's text or against the majority reading. Where the readings decide, they add to the edition: the map joins S (Schaffhausen, Min.~75), which Tremp set apart, to the Reims pair L2 B4 on 26 readings, all in S's main text and uncorrected. The two counts root the tree differently: against Tremp's text the root lies on W's own edge, his top division; against the majority it lies between W, T, P and the other four. The choice is the editor's, and Tremp states it: two passages found only in W, P and T he judges authentic, so that the other branch omitted them~\citep[42--43]{Tremp1995}. With that judgement W, T and P share only original readings, which by Lachmann's rule prove nothing; with the opposite one the same places make them a family. The readings fix the shape of the tree; its direction rests on one judgement about two passages, and the map names them.

Where the readings leave the root open, the decision falls to the editor, and here it rests on a judgement of authenticity that a reader can weigh, because Tremp states it and the map shows that nothing else decides. This is the division the section has traced across the \emph{Monumenta}: the lower families come out of the editors' own apparatus, and the map adds to them where the readings decide, as with S; the top division, and with it the root, rests on judgements the apparatus does not record. An edition that states those judgements, as Tremp's does, lets its stemma be checked to the end.

\begin{figure*}[t]
\centering
\includegraphics[width=0.78\textwidth]{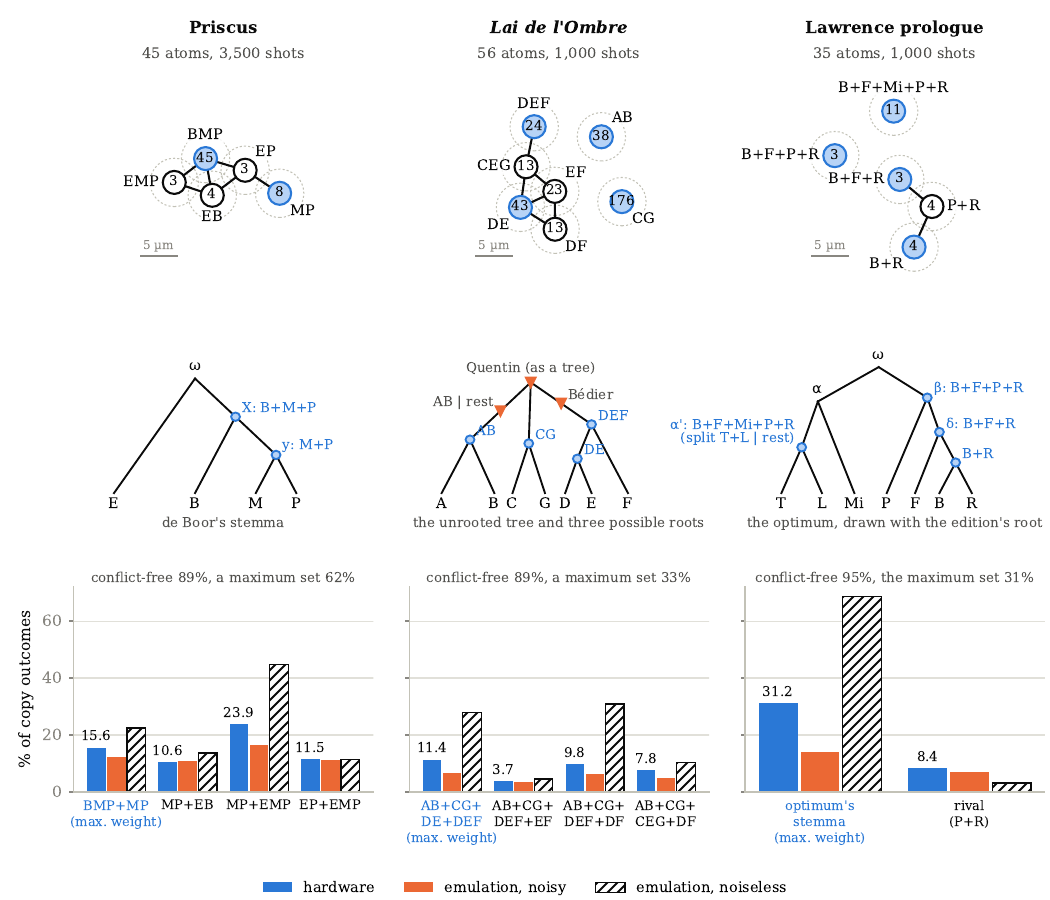}
\caption{\label{fig:priscus}Three stemmata computed on a neutral-atom quantum processor, Pasqal's Ruby at TGCC (GENCI): Priscus, the \emph{Lai de l'Ombre} and the Lawrence prologue. Top: one copy of each register as placed on the device. Each atom is a candidate family, labelled with its support (shared departures from the printed text; for Lawrence, units supporting the split, oriented away from T, so that B+F+Mi+P+R is the split of T+L from the rest; for the \emph{Lai}, DE carries 43, from an earlier extraction, rather than 41); an edge joins two families that cannot both hold; dashed disks have radius $R_b/2$, so atoms whose disks overlap block each other; shaded atoms form the maximum-weight set. Middle: the corresponding stemmata, with three rootings of the \emph{Lai}: next to DEF (Bédier), at the centre, and next to AB. Bottom: each run returns sets of families, one per outcome; the bars give the share of outcomes that are each largest possible set of families with no two crossing (for Lawrence, the optimum and its rival), on hardware and in emulation with the device's nominal noise. Above each panel: the share of outcomes in which no two kept families cross, so that they form a stemma, and the share that are one of these largest sets.}
\end{figure*}

%=====================================================================
\section{Stemmata on neutral-atom quantum processors}
\label{sec:quantum}
%=====================================================================

\subsection{A physical solver for the stemma}
\label{sec:native}

All the instances above are solved exactly on an ordinary computer, most in seconds. Exactness has a cost that grows with the tradition: on the benchmark with known history, proving optimality takes about a minute with eight solver threads and was not reached in three hours with one, and traditions with thousands of candidate families are where exact optimisation becomes expensive. At that scale other hardware is worth testing, and one kind of quantum processor solves the problem of the stemma natively.

Quantum processors have emerged over the last decade as specialised machines, closer to accelerators than to general-purpose computers, with the prospect of solving certain tasks faster or at lower energy cost than conventional computers~\citep{JaschkeMontangero2023}; combinatorial optimisation, measured by the time to reach an optimal answer, is among the tasks most often proposed~\citep{Dalyac2024}. One family of such processors traps single neutral atoms in ultra-high vacuum, each in the focus of a laser beam, in planar arrays of chosen geometry: commercially accessible machines offer about one hundred to a few hundred atoms, and defect-free arrays of about a thousand have been demonstrated~\citep{Lim2026}. Used in analog mode, without the digital gates of a quantum circuit, they encode a graph problem directly in the positions of the atoms and in the laser drive~\citep{Henriet2020,Dalyac2024}.

The stemma maps onto such a processor term by term. Each candidate family is one atom, and exciting the atom to its Rydberg state means keeping the family. Two atoms closer than the blockade radius $R_b$ (here about 7~$\mu$m) cannot both be excited, so placing two crossing families within $R_b$ makes keeping both impossible in the ideal blockade model (on the device, strongly suppressed): the rule that crossing families exclude each other is enforced by the physics of the device rather than computed. Compatible families are placed farther apart. Nothing is translated into penalty terms and no auxiliary atom is needed; the one condition is geometric, that the conflict graph be a \emph{unit-disk graph}, one whose vertices can be placed in the plane so that two are joined exactly when they lie within a fixed distance of each other.

The support of each family enters through the laser detuning, which sets how much each excitation lowers the energy of the array. With the same detuning on every atom, the lowest-energy configuration is the largest set of families without conflict, a maximum independent set, whatever their support. The best-supported stemma, the maximum-weight independent set of Eq.~\eqref{eq:mwis}, needs a detuning set atom by atom, in proportion to the support of each family; with it, the stemma is the ground state of the device in the ideal model. This per-atom setting is what turns the processor from a solver of the unweighted problem into a solver of the stemma, and the processor used below does not yet offer it.

The machine reaches the ground state by sweeping the detuning from negative to positive values over a few microseconds, after which the atoms are imaged. Each run, or shot, returns one configuration, one candidate stemma, and returns the optimum only with some probability: the processor is used statistically, and its performance is the probability that a single shot is optimal.

\subsection{Three small instances on the quantum processor}
\label{sec:ruby}
We ran three instances on Ruby, Pasqal's neutral-atom processor of up to 100 atoms at TGCC (GENCI), with one drive fixed in advance and several copies of each instance in one register, so that every shot returns several outcomes (Fig.~\ref{fig:priscus}; settings in Appendix~\ref{app:quantum}). The first is the fifth-century historian Priscus of Panium, whose \emph{History} survives in fragments, most of them in the \emph{Excerpta de legationibus}, edited by Carl de Boor~\citep{deBoor1903}: the apparatus of four witnesses yields five candidate families, and the maximum-weight set is de Boor's stemma exactly, while four sets tie without weights. The second is the \emph{Lai de l'Ombre} (Sec.~\ref{sec:lai}), restricted to its well-supported families, with the supports of an earlier extraction (DE 43 rather than 41), whose maximum-weight set is the unrooted tree found above; four sets again tie without weights. The third is the core of the Lawrence stemma (Fig.~\ref{fig:lawrence}), whose unique maximum set is the optimum's stemma.

On all three the device does what the encoding asks: most outcomes keep no two crossing families together (89 to 95\%), and the maximum sets come out more often (62, 33 and 31\% of outcomes) than emulation with the device's nominal noise predicts (51, 21 and 14\%); the best-supported stemma itself comes out in 16, 11 and 31\%, and the remaining outcomes are smaller sets or, in 5 to 11\%, sets with a conflict. On instances this small, a classical greedy choice of compatible families, in random order, finds a maximum set with probability 1, 1 and 2/3. Without weights the device cannot prefer the best-supported of the stemmata the readings allow.

A detuning set atom by atom changes this. Ruby does not yet offer it, so it was tested in emulation, on the same drive, with each atom's detuning in proportion to the support of its family (with a floor of 0.3). On Priscus, weights raise the probability of de Boor's stemma from 0.22 to 0.57 on an ideal device, and from 0.12 to 0.19 with the device's nominal noise. With slower sweeps, on layouts checked against the weak interactions between compatible families (new ones for the \emph{Lai} and Lawrence), the best-supported stemma becomes more frequent on all three instances, and the other stemmata come out largely in the order of their support. A further simulation asks whether the device could return the competing stemmata in proportion to their evidence. With the weights scaled by the uncertainty of the evidence and the device calibrated for its setting and layout, on sixteen test cases drawn from the Lawrence and \emph{Lai} collations and from Priscus, an ideal device gives the leading stemma of each case within 7 percentage points, on average, of how often resampling the readings favours it.

The instances that fit today are small and easy on an ordinary computer, and no speed-up is claimed. They show that the encoding is faithful, with maximum sets at least as frequent as the noise model predicts, and they give the hardware problems drawn from real evidence, with a known answer, in which the unweighted and weighted problems differ. Traditions of editorial interest need larger registers with per-atom detunings.

\subsection{Larger traditions}
\label{sec:fitting}

A graph that is not unit-disk can be placed on the array with embedding gadgets, chains of auxiliary atoms that copy an atom's state to where its conflicts are~\citep{Dalyac2024}, but at our scale the cost is prohibitive. A nine-witness tradition collated by machine, the \emph{In visionem Ezechielis} of the twelfth-century theologian Richard of St Victor, has 15 to 47 candidate families depending on the support threshold; embedded with gadgets, these graphs would need 524 to 6{,}710 atoms. Stemmatic data allow the instance to be made native instead, in four ways:
\begin{enumerate}
\item Thresholding the support: keep the families supported by at least $s$ units, with $s$ the smallest value that admits a unit-disk layout. For the Lawrence prologue, $s=3$ leaves five families and two conflicts, laid out on Ruby with conflicting pairs at $0.92\,R_b$ or less and compatible pairs at $1.42\,R_b$ or more.
\item Encoding one branch at a time, where a supported backbone is available: the interesting conflicts then sit inside branches. In the Donatus tradition, the families inside $\upsilon$ and $\tau$ supported by at least five units form a twelve-atom native graph, precisely where the L5 question arises (Sec.~\ref{sec:evidencemap}); at four units we found no unit-disk layout.
\item Compressing the weights: supports $w_i$ become detunings $\Delta_i=\Delta\,(a+(1-a)\,w_i/w_{\max})$, with $\Delta$ the largest detuning the device allows and $a<1$ a floor. Compression adds the same reward for every family kept, so when competing sets differ in size it can change which set is best; each compressed instance must be checked against the original optimum, and the layout of the atoms checked so that the weak interactions between compatible families do not change it either.
\item Tiling copies: a small instance fits many times in one register (nine Priscus copies in 45 atoms, seven Lawrence copies in 35), so each shot yields several outcomes, correlated through the shared shot and device conditions.
\end{enumerate}
On the larger Lawrence graph of the twelve families supported by at least two units ($s=2$), the weights matter even more than for Priscus: its maximum independent set is nine-fold degenerate, while the maximum-weight set is unique and is the optimum of Fig.~\ref{fig:lawrence}, the groups of the stemma drawn by hand with B and R joined inside $\delta$.

%=====================================================================
\section{Discussion and conclusions}
\label{sec:conclusion}
%=====================================================================

Stemma reconstruction is a maximum-weight independent set on a conflict graph. In this form the problem can be solved exactly, fragmentary witnesses enter through joint completion instead of being dropped, and every family of the result comes with its reasons: the units that support it, the better-supported families that exclude its rivals, and its stability under resampling. Exactness certifies the optimum of the stated objective, not a better reconstruction: on a tradition with known history the method stands level with maximum parsimony and Bayesian inference, which make largely the same errors on the same readings.

Its value is that every result can be traced to the readings. On Donatus the evidence map separates stable from fragile families and leads back to the loci that decide them; on the five \emph{Monumenta} case studies it recovers the editors' groups and, where it departs from them, lands on questions they had raised or decided on evidence outside the apparatus. Across the series the top division is often not settled by the variants printed in the apparatus. Generative models propose good trees, but the evidence they cite for them is unreliable.

The method separates what the old debate disputed into steps that can each be checked~\citep{Roelli2020,Trovato2017}: the groups, taken from all shared readings as Quentin proposed; their orientation, which needs a stated judgement on errors, as the genealogical method requires; and the limits of the evidence, where Bédier located the editor's freedom. In the \emph{Lai de l'Ombre} and in Thegan the tree is fixed and the root rests on a judgement about which readings are original; in the Lawrence prologue a near tie inside $\delta$ is decided by reading R's agreements with P as correction from a second exemplar.

The mathematics also gives Bédier's suspicion a ground. When units are independent and identically distributed, the method recovers the true stemma as the text grows if the true families are in expectation the unique best-supported compatible set, and fails when another compatible set has strictly greater expected support, as in the textual form of Felsenstein's long-branch zone~\citep{Felsenstein1978}. A \emph{polytomy}, a lost exemplar copied independently three times or more, is such a case. A method that keeps the best-supported tree tends to turn a true three-way branching into a two-way one: a reading that two of three independent copies share, by coincident error or because the third alone changed the text, proposes a family that crosses no true one; the best-supported tree adds it, and as the text grows such an agreement becomes almost certain to occur~\citep{JurczakCoincidence}. The mechanism is Timpanaro's ``the majority becomes a totality''~\citep[180--182]{Timpanaro2005}, which Haugen~\citep[606--608]{Haugen2016} and Spyridonova and Kurbanov~\citep{Spyridonova2023} trace to the common-error method itself. It adds to the explanation by loss~\citep{Weitzman1987,GuidiTrovato2004,Hoenen2017}, which a calibrated birth--death model has made precise: 65\% of the stemmata simulated under its constant-rate fit are two-branched at the root, against 77\% of 117 editors' stemmata of Old French romances~\citep{Camps2025}. Simulated on the same model, best-supported reconstruction that admits a family on two to five shared departures reproduces the editors' root share; if editors grouped manuscripts as that estimator does, a working hypothesis, over-resolution is a plausible contributor to the excess, though not its whole explanation; it is the textual form of the star-tree paradox of phylogenetics~\citep{Lewis2005,SteelMatsen2007}, seen also on an artificial tradition~\citep{Spencer2004}. The remedy is to require each family to clear a level of support above what chance agreements produce, measured on the collation itself.

These results have limits. An apparatus is selective~\citep[37--38, 168]{Roelli2020}, so a test on the apparatus tests the edition's evidence, not the manuscripts', and both the case studies and the series test measure consistency with each editor's evidence, not independence from it. On the benchmark with known history several solutions are optimal, and the method resolves more families than the true tree has. The Donatus matrix comes from two AI transcriptions of a printed apparatus, reconciled on the page images and spot-checked against three witnesses; the \emph{Monumenta} case studies come from OCR, and the series test from vision-model transcriptions; the reference transcriptions of the test and proof pages are AI-made and were not checked by a human reader; a blind check of 60 extracted departures on the page images, 40 of them behind the four contradicted divisions, found every one recorded with the right manuscripts, and the readings on which the case studies turn were read in the witnesses themselves. The coding of readings matters, and contamination, which Maas held beyond remedy in 1927~\citep[49]{Maas1958} and the classical scholar Martin West called in 1973 ``a normal state of affairs''~\citep[5]{West1973}, is flagged by the map but not yet modelled. On hardware, the quantum part is a proof of concept on small instances, for the unweighted problem only; the weighted problem is shown in emulation.

An edition's apparatus is a dataset. For editors, the evidence map says before a new edition which branches of the existing stemma are secure and which few loci, in which manuscripts, would decide the others; in the edition itself each branch can be printed with its evidence. For the field, the critical editions of the nineteenth and twentieth centuries form a corpus of stemmata that can now be audited series by series (the \emph{Monumenta}, the \emph{Corpus Scriptorum Ecclesiasticorum Latinorum}, the \emph{Corpus Christianorum}, the Teubner texts), at low cost, and a flagged top is a forecast that a future edition can confirm or refute. For quantum hardware, stemmata are real problems whose answers an expert can read; arrays of about a thousand atoms~\citep{Lim2026}, given per-atom detunings, might reach thresholded instances of editorial interest, provided thresholding is checked not to change the optimum of each instance. Such a device could do more than solve: if it returned the competing stemmata in proportion to the strength of their evidence, it would show the uncertainty of a stemma rather than a single tree, which the simulations of Sec.~\ref{sec:ruby} suggest is within reach of an ideal device.

Textual criticism remains an art as well as a science; the method does not replace the editor's judgement, but marks where the science ends and that judgement begins. We hope the method, its code and the data released with this paper will be of use to editors of every school described above: to those who build stemmata, as a way to state the evidence behind each branch, and to those who doubt them, as a way to show where that evidence runs out.

\section*{Acknowledgements}
The author thanks Don Federico Corrubolo and the Archivio Capitolare Lateranense, Rome, for access to A.~81, the Biblioteca Ambrosiana, Milan, for access to B~55~inf., and Dr Gaia Elisabetta Unfer Verre and the Archivio Storico Diocesano di Lucca for photographs of its passionals P+ and B. This access and these images were provided for the author's work on the prologue \emph{In divinis laudibus}. The bifolium MC-C-077 (siglum L) belongs to the author's Metamorphoses of Civilization Collection, Dallas, which also holds the photographs of it reproduced in Fig.~\ref{fig:plates}a, b. 

\section*{Data availability}
The mathematics, with complete proofs, machine checks and a formalisation in Lean, is in the technical report~\citep{JurczakReport}. The code, the data and all derived results are in the code and data deposit~\citep{StemmaMWIS}, which also releases the critical apparatus of the 42 \emph{Auctores antiquissimi} texts located for the series test as tables: 150{,}947 readings, each with its witnesses, page image and apparatus line (CC BY 4.0). The Heinrichi data are distributed by the challenge organisers~\citep{CASC2007}. The Donatus matrix is derived from a copyrighted apparatus and is not redistributed; the code rebuilds it from a transcription of the printed edition.

\section*{AI disclosure statement}
AI tools were used under the author's direction. Claude (Anthropic, Claude Opus 5.5, through Claude Code) wrote code; an AI mathematics assistant (Theo, FirstPrinciples) helped develop the formal statements. AI agents transcribed the Donatus apparatus and recorded the editors' groupings from their prefaces, and a vision model (Gemini, \texttt{google/gemini-4-argon}) transcribed the 4{,}091 \emph{Monumenta} pages; beyond the checks reported in Secs.~\ref{sec:experiments} and~\ref{sec:bedier}, these transcriptions were not checked by a human reader. The author checked the proofs, the reported numbers and the manuscript readings. The language models compared in Sec.~\ref{sec:llm} are objects of study, accessed through OpenRouter. 

\section*{Funding}
This research received no external grant funding. This work was granted access to the high-performance computing resources of the Très Grand Centre de Calcul (TGCC), including the Pasqal Ruby processor, under the allocation AD010617947 (project iQON) made by GENCI (Grand Équipement National de Calcul Intensif).

\section*{Competing interests}
The author holds shares in Pasqal, the manufacturer of the Ruby processor. Access to Ruby was obtained through the standard GENCI allocation process.

\bibliographystyle{agsm}
\bibliography{refs}

\clearpage
\onecolumn
\section*{Appendices}
\appendix
\section{Texts, editions and witnesses}
\label{app:witnesses}

Table~\ref{tab:editions} lists every tradition used in this paper, with the edition it is read from and what the method finds. Unless stated otherwise the data are the edition's apparatus, not a fresh collation, and sigla are the editor's.

\begin{table}[H]
\caption{\label{tab:editions}\label{tab:texts}The traditions used in this paper. Edition: the edition whose apparatus is read. Witnesses: number of witnesses, with the editor's grouping where relevant. Used in: the section that uses the tradition. Finding: what the method finds. Shelfmarks of the Getica, Priscus and Thegan witnesses are in the code and data deposit (see Data availability).}
\scriptsize\setlength{\tabcolsep}{3pt}
\begin{ruledtabular}
\begin{tabular}{p{0.20\textwidth}p{0.20\textwidth}p{0.14\textwidth}p{0.14\textwidth}p{0.24\textwidth}}
\toprule
\rr{Text} & \rr{Edition} & \rr{Witnesses} & \rr{Used in} & \rr{Finding} \\
\midrule
\rr{\emph{Heinrichi}, Finnish text copied by volunteers (artificial)} & \rr{Computer-assisted stemmatology challenge, 2007~\citep{RoosHeikkila2009,CASC2007}} & \rr{37, true history known} & \rr{benchmark (Sec.~\ref{sec:experiments}, \ref{sec:llm})} & \rr{exact optima (79.1--82.3) level with parsimony and Bayesian inference; best language model 83.3} \\
\rr{\emph{Passio} of Donatus of Arezzo (BHL 2294)} & \rr{\citet{Licciardello2018}} & \rr{54 manuscripts; 12 named families} & \rr{benchmark, evidence map (Sec.~\ref{sec:experiments}, \ref{sec:evidencemap})} & \rr{8--10 of the editor's 12 families (exact optima); top family $\nu$ has no supporting reading; placement of L5} \\
\rr{Prologue \emph{In divinis laudibus} to the passion of St Lawrence (BHL 4759), 11th--12th c.} & \rr{this study; reading text online} & \rr{7: T, L, F, P, B, R, Mi (Table~\ref{tab:sigla})} & \rr{new stemma; quantum instance (Sec.~\ref{sec:lawrence}, \ref{sec:ruby})} & \rr{the groups of the stemma drawn by hand; B+R and P+R a near tie (0.42, 0.45)} \\
\rr{Priscus of Panium (5th c.), in the \emph{Excerpta de legationibus}} & \rr{\citet{deBoor1903}} & \rr{4 (E, B, M, P), 16th-c.\ copies} & \rr{quantum instance (Sec.~\ref{sec:ruby})} & \rr{maximum-weight set = de Boor's stemma; four tied sets without weights} \\
\rr{Jordanes, \emph{Getica} (551)} & \rr{Mommsen, AA~5,1, 1882~\citep{Mommsen1882}; revised by Giunta and Grillone 1991~\citep{GiuntaGrillone1991}, Grillone 2017~\citep{Grillone2017}} & \rr{11: H, P, V, L, A (Ambrosianus), S (epitome), O, B, X, Y, Z} & \rr{case study (Sec.~\ref{sec:mommsen})} & \rr{three orders recovered; A of mixed affiliation; no ancestor of orders I and II, as in Grillone 2017} \\
\rr{Eutropius, \emph{Breviarium}} & \rr{Droysen, AA~2, 1879~\citep{Droysen1879}} & \rr{Table~\ref{tab:mgh}} & \rr{case study (Sec.~\ref{sec:mgh})} & \rr{Droysen's families; the agreements of classes B and C disputed with Mommsen} \\
\rr{Venantius Fortunatus, poems (6th c.)} & \rr{Leo, AA~4,1, 1881~\citep{Leo1881}} & \rr{Table~\ref{tab:mgh}} & \rr{case study (Sec.~\ref{sec:mgh})} & \rr{Leo's interpolated pairs; 55 places printed against the best manuscripts} \\
\rr{Sidonius Apollinaris, letters and poems (5th c.)} & \rr{Luetjohann, filiation by Leo, AA~8, 1887~\citep{Luetjohann1887}} & \rr{Table~\ref{tab:mgh}} & \rr{case study (Sec.~\ref{sec:mgh})} & \rr{M sides with CFP, not with T, on spellings} \\
\rr{Cassiodorus, \emph{Variae} (6th c.)} & \rr{Mommsen, AA~12, 1894~\citep{Mommsen1894}} & \rr{Table~\ref{tab:mgh}} & \rr{case study (Sec.~\ref{sec:mgh})} & \rr{Mommsen's classes; class II not one family} \\
\rr{Jean Renart, \emph{Lai de l'Ombre} (early 13th c.), about 960 lines} & \rr{Bédier 1890; 2nd ed.\ 1913~\citep{Bedier1913}, read here} & \rr{7 (A--G), Bibliothèque nationale} & \rr{Bédier's case (Sec.~\ref{sec:lai}); quantum instance (Sec.~\ref{sec:ruby})} & \rr{tree ((A,B),(C,G),((D,E),F)) fixed; root left open} \\
\rr{Thegan, \emph{Gesta Hludowici imperatoris} (9th c.)} & \rr{Tremp, MGH SRG 64, 1995~\citep{Tremp1995}} & \rr{12, 7 complete} & \rr{case study (Sec.~\ref{sec:thegan})} & \rr{tree fixed at stability 1.00; root follows Tremp's authenticity judgement; S joins Reims} \\
\rr{Richard of St Victor, \emph{In visionem Ezechielis} (12th c.)} & \rr{our machine collation} & \rr{9} & \rr{large instance (Sec.~\ref{sec:fitting})} & \rr{15 to 47 candidate families, beyond today's processors} \\
\bottomrule
\end{tabular}
\end{ruledtabular}
\end{table}

All seven copies of the prologue were collated in full (Table~\ref{tab:sigla}): T, F, P and B on reproductions, L on photographs taken with the folds lifted, R and Mi on the manuscripts, read in person on 3 and 5 November 2026 (photography is not allowed at the Ambrosiana). T and L are recorded neither in the BHL nor in the census of Latin legendaries of the Institut de recherche et d'histoire des textes (IRHT). All but F have the long ending, to Lawrence's arrest. L carries on its first leaf the end of the passion of Donatus and on its second the prologue; T has, on consecutive pages, the passion of Donatus, the whole prologue and the passion of Lawrence; R and Mi have the prologue after Donatus, Cyriacus and Romanus.

\begin{table}[H]
\caption{\label{tab:sigla}The seven manuscripts of the prologue \emph{In divinis laudibus} (BHL 4759), with their sigla, dates and the part of the text each contains.}
\begin{ruledtabular}
\footnotesize\setlength{\tabcolsep}{3pt}
\begin{tabular}{lp{0.55\linewidth}p{0.32\linewidth}}
\toprule
Siglum & Manuscript & Date; extent \\
\midrule
T & Trento, Biblioteca comunale, BCT1-1566, pp.~124a--126a & 12th c., 2nd half; complete \\
L & Dallas, TX, Metamorphoses of Civilization Collection, MC-C-077, fols 2r--2v & 12th c., 2nd half; from \emph{Quique per spiritum sanctum} \\
F & Florence, Biblioteca Medicea Laurenziana, Plut.~20.3, fols 144va--145ra & 12th c., Lucca; to \emph{Agapito} \\
P & Lucca, Biblioteca Capitolare Feliniana, Passionario P+, fols 47r--47v & 12th c., 1st half; complete \\
B & Lucca, Biblioteca Capitolare Feliniana, Passionario B, fols 234v--235r & 12th c., last quarter; complete \\
R & Rome, Archivio Capitolare Lateranense, A.~81, fols 54v--55r & 12th c., 1st quarter, Lucca; complete \\
Mi & Milan, Biblioteca Ambrosiana, B~55 inf., fols 63r--63v & 11th or 12th c.; complete \\
\bottomrule
\end{tabular}
\end{ruledtabular}
\end{table}

\section{The series test}
\label{app:series}

\begin{table}[H]
\caption{\label{tab:seriestexts}The fourteen texts of the series test, with the editor's top division and the verdict. Edition: editor and volume of the \emph{Auctores antiquissimi}. Witnesses: number of witnesses analysed and the editor's top division, which may leave some of them unplaced (A for Victor; eight of the fourteen for Prosper, C among them). Finding: the verdict, with the families that decide it and their weights in units.}
\begin{ruledtabular}
\begin{tabular}{p{0.24\textwidth}p{0.24\textwidth}p{0.20\textwidth}p{0.26\textwidth}}
\toprule
\rr{Text} & \rr{Edition} & \rr{Witnesses} & \rr{Finding} \\
\midrule

\rr{Eugippius, \emph{Vita Severini} (511)} & \rr{Sauppe, AA~1,2 (1877)} & \rr{4; top L $|$ VMD} & \rr{supported (VM 214, VMD 51)} \\
\rr{Victor of Vita, \emph{Historia persecutionis} (late 5th c.)} & \rr{Halm, AA~3,1 (1879)} & \rr{5; top BLV $|$ R} & \rr{supported (BLV 149)} \\
\rr{Sidonius, \emph{carmina} (5th c.)} & \rr{Luetjohann, filiation by Leo, AA~8 (1887)} & \rr{5; top M $|$ TCFP} & \rr{supported (TCFP 32)} \\
\rr{Marcellinus Comes, \emph{Chronicon} (6th c.)} & \rr{Mommsen, AA~11 (1894)} & \rr{4; top S $|$ TUR} & \rr{supported (TUR 43)} \\
\rr{Ennodius, works (early 6th c.)} & \rr{Vogel, AA~7 (1885)} & \rr{7; top B $|$ VLTCOP} & \rr{contradicted (BVL 211)} \\
\rr{Fortunatus, \emph{Vita Martini} (6th c.)} & \rr{Leo, AA~4,1 (1881)} & \rr{8; top PTF $|$ NBLMS} & \rr{contradicted (PN 75)} \\
\rr{Paul the Deacon, \emph{Historia Romana} (8th c.)} & \rr{Droysen, AA~2 (1879)} & \rr{6; top HVBM $|$ NZ} & \rr{contradicted (NZM 48)} \\
\rr{Prosper, \emph{Epitoma chronicon} (5th c.)} & \rr{Mommsen, AA~9 (1892)} & \rr{14; top AORV $|$ MY} & \rr{contradicted (AC 108)} \\
\rr{Claudian, \emph{carmina maiora} (c.~400)} & \rr{Birt, AA~10 (1892)} & \rr{9; top VPGM $|$ CBAJR} & \rr{silent (VP 362, BA 191)} \\
\rr{Claudian, \emph{De raptu Proserpinae}} & \rr{Birt, AA~10 (1892)} & \rr{8; top FVS $|$ CABDW} & \rr{silent (CB 22)} \\
\rr{\emph{Historia Brittonum} (9th c.)} & \rr{Mommsen, AA~13 (1898)} & \rr{11; top HKPQDCGL $|$ ZMN} & \rr{silent (MN 686)} \\
\rr{Isidore, \emph{Chronica maiora} (7th c.)} & \rr{Mommsen, AA~11 (1894)} & \rr{19; top 12 $|$ 7 witnesses} & \rr{silent (low pairs only)} \\
\rr{Isidore, \emph{Historia Gothorum} (7th c.)} & \rr{Mommsen, AA~11 (1894)} & \rr{10; top P $|$ BNHOXYRMA} & \rr{silent (MA 80)} \\
\rr{Jordanes, \emph{Romana} (551)} & \rr{Mommsen, AA~5,1 (1882)} & \rr{4; one branch, HPVL, stated} & \rr{silent (HPV 136)} \\
\bottomrule
\end{tabular}
\end{ruledtabular}
\end{table}

The admission of texts, the reasons for the thresholds, the preregistration, the robustness checks and the count of two-way tops are in the code and data deposit (see Data availability).

\section{Generative language models}
\label{app:llm}

Every siglum was replaced by an anonymous label in a fixed random order, which removes the obvious cues for recognising a tradition but cannot rule out recognition from the readings themselves; no permuted or freshly simulated tradition was run as a control. Every model received the same collation matrix, with the task of returning nested families in a fixed format, and its families were scored exactly as for MWIS. Each model was run three times at its highest reasoning setting through one routing service. Table~\ref{tab:llm} gives the results; for Claude Fable 5.2, three further runs whose answers were not kept scored 83.6, 84.3 and 77.3 on Heinrichi (mean over all six runs 82.5) and are not in the table. Table~\ref{tab:llmev} checks, by program against the matrix, every piece of evidence the models cite: 306 claims in 24 answers from the eight models that cite checkable evidence. Of these, 192 (63\%) hold for every witness named, 92 (30\%) for only part of the group, and 16 are wrong; of the 192, only 125 are shared by no witness outside the group, which is what a family needs.

\begin{table}[H]
\caption{\label{tab:llm}Stemmata proposed by generative language models, against exact MWIS and existing methods (in italics). Each model received the collation with every siglum replaced by an anonymous label and was asked for the families of the stemma. Heinrichi: skeleton score against the true tree (\%), mean and range over the runs that gave a usable answer. Donatus: the editor's families recovered exactly (of 12). Answers: usable answers of the three requested per text. Tokens: median output tokens per request, in thousands, failed requests included.}
\begin{ruledtabular}
\begin{tabular}{lccccc}
\toprule
Method & Heinrichi (\%) & Donatus (of 12) & Answers & Tokens (k) \\
\midrule
\emph{MWIS, exact} & \emph{79.1--82.3} & \emph{8--10} & & \\
\emph{Hybrid, $\theta=0.8$--$0.9$} & \emph{81.7} & \emph{10} & & \\
\emph{Maximum parsimony} & \emph{81.7} & \emph{9} & & \\
\emph{Bayesian (MrBayes)} & \emph{82.2}$^{a}$ & \emph{8} & & \\
\emph{Neighbor-Net, greedy tree} & \emph{71.8} & \emph{8} & & \\
\emph{Neighbour joining} & \emph{67.9} & \emph{8} & & \\
\midrule
Claude Fable 5.2 & \textbf{83.3} (82.0--84.2) & 8.7 (8--9) & 6/6 & 92 \\
GPT-6.1 Sol & 72.0 (69.2--75.7) & 7.7 (7--8) & 6/6 & 48 \\
Claude Opus 5.5 & 68.0 & 7.3 (6--8) & 4/6 & 9 \\
Kimi K3.1 Pro & 65.0 (62.2--66.4) & 6.7 (6--7) & 6/6 & 62 \\
Claude Sonnet 5.5 & 62.8 (61.0--64.6) & 6.0 & 6/6 & 3 \\
DeepSeek V4.1 Flash & 59.6 (56.1--63.0) & 4.0 & 4/6 & 96 \\
Grok 4.7 & 57.7 (53.5--65.3) & 3.3 (3--4) & 6/6 & 83 \\
Mistral Large 4 & 56.7 (52.1--62.1) & 4.0 & 6/6 & 51 \\
MiniMax M3 & 56.4 & 1.0 & 2/6 & 124 \\
Qwen 3.8 Max & 54.4 (52.5--56.1) & 5.0 & 5/6 & 111 \\
Nemotron 3 Ultra & 52.0 (49.3--54.5) & 3.0 & 6/6 & 37 \\
GLM 5.3 Prime & --- & 5.0 (4--7) & 3/6 & 126 \\
Gemini 4 Argon, Qwen 3.8 27B & --- & --- & 0/6 & 41--66 \\
Command A+ & --- & 0.0 & 1/6 & 5 \\
\bottomrule
\end{tabular}
\end{ruledtabular}
{\raggedright\footnotesize $^{a}$Runs not converged; consensus stable (Table~\ref{tab:results}). Failures are the model's own output limit while reasoning (Gemini, MiniMax), empty or unparsable answers, and refusals (Opus on Heinrichi).\par}
\end{table}

\begin{table}[H]
\caption{\label{tab:llmev}Whether the evidence that language models cite for their families is true, checked against the collation they were given (24 answers from eight models). Claims: readings, lacunae or extra passages said to be shared by named witnesses. Holds: true of every witness named. Distinct.: of those, shared by no other witness, as a family requires. Part: true of only some of the witnesses named. Wrong: true of none of the witnesses named. Six further claims are true with the place off by up to one block (2) or are said to be exclusive but are shared by others (4). Kimi, Qwen, DeepSeek and MiniMax made no checkable claim.}
\begin{ruledtabular}
\setlength{\tabcolsep}{3pt}
\begin{tabular}{lrrrrr}
\toprule
Model & Claims & Holds & Distinct. & Part & Wrong \\
\midrule
Opus 5.5 & 95 & 58 & 47 & 37 & 0 \\
Sonnet 5.5 & 68 & 27 & 15 & 39 & 2 \\
Fable 5.2 & 34 & 25 & 5 & 4 & 3 \\
GLM 5.3 & 34 & 24 & 14 & 5 & 5 \\
Grok 4.7 & 31 & 28 & 28 & 0 & 0 \\
Nemotron 3 & 25 & 19 & 7 & 5 & 0 \\
Mistral Large 4 & 16 & 9 & 7 & 2 & 5 \\
GPT-6.1 Sol & 3 & 2 & 2 & 0 & 1 \\
\midrule
All & 306 & 192 & 125 & 92 & 16 \\
\bottomrule
\end{tabular}
\end{ruledtabular}
\end{table}

\section{Quantum runs}
\label{app:quantum}

The runs used Ruby (HPCQS1), the analogue neutral-atom quantum computer built by Pasqal for up to 100 rubidium atoms and installed at TGCC (GENCI) within the EuroHPC project HPCQS. One drive, fixed in advance (a 6~$\mu$s detuning sweep at constant Rabi frequency, no per-atom detuning), served all three instances, each tiled several times in one register: Priscus nine times in 45 atoms (3{,}500 shots), the \emph{Lai} eight times in 56 atoms (1{,}000 shots), Lawrence seven times in 35 atoms (1{,}000 shots). Emulations use Pulser~\citep{Silverio2022pulser} with its QuTiP state-vector emulator, ideal or with the device's nominal noise (readout false positives 2\%, false negatives 10\%, dephasing time 4.5~$\mu$s). Registers, raw results, emulations and the scripts that score them are in the code and data deposit (see Data availability).

\end{document}